%% file: main.tex
\documentclass[a4paper,11pt]{article}
\usepackage{jheppub}
\usepackage{amsmath} 
\usepackage{amsthm} 
\usepackage{amssymb}	
\usepackage{graphicx} 
\usepackage{slashed}
\usepackage{hyperref}
\usepackage{breqn}
\usepackage{feynarts}
\allowdisplaybreaks

\def\Feynarts{{{\sc FeynArts}}}
\def\Feyncalc{{{\sc FeynCalc}}}
\def\mathematica{{{\sc mathematica}}}

\newcommand{\nn}{\pagebreak[0] \nonumber \\}

\usepackage{subfig}
\usepackage{tikz}
\usetikzlibrary{"arrows", "automata", "backgrounds", "calendar", "chains", "matrix", "mindmap", "patterns", "petri", "shadows", "shapes.geometric", "shapes.misc", "spy", "trees"}
\usetikzlibrary{arrows,shapes}
\usetikzlibrary{trees}
\usetikzlibrary{matrix,arrows} 	
\usetikzlibrary{positioning}				
\usetikzlibrary{calc,through}				
\usetikzlibrary{decorations.pathreplacing}
\usepackage{pgffor}

\usetikzlibrary{decorations.pathmorphing}	
\usetikzlibrary{decorations.markings}
\tikzset{
    vector/.style={decorate, decoration={snake}, draw},
	provector/.style={decorate, decoration={snake,amplitude=2.5pt}, draw},
	antivector/.style={decorate, decoration={snake,amplitude=-2.5pt}, draw},
    fermion/.style={draw=black, postaction={decorate},
        decoration={markings,mark=at position .55 with {\arrow[draw=black]{>}}}},
    fermionbar/.style={draw=black, postaction={decorate},
        decoration={markings,mark=at position .55 with {\arrow[draw=black]{<}}}},
    fermionnoarrow/.style={draw=black},
    gluon/.style={decorate, draw=black,
        decoration={coil,amplitude=1.5pt, segment length=3pt}},
    scalar/.style={dashed,draw=black, postaction={decorate}},
    scalarbar/.style={dashed,draw=black, postaction={decorate},
        decoration={markings,mark=at position .55 with {\arrow[draw=black]{<}}}},
    scalarnoarrow/.style={dashed,draw=black},
    electron/.style={draw=black, postaction={decorate},
        decoration={markings,mark=at position .35 with {\arrow[draw=black]{>}}}},
    positron/.style={draw=black, postaction={decorate},
        decoration={markings,mark=at position .35 with {\arrow[draw=black]{<}}}},        
	bigvector/.style={decorate, decoration={snake,amplitude=4pt}, draw},
}
\tikzstyle{block} = [draw, rectangle, 
    minimum height=3em, minimum width=6em]
    
\newcommand{\semiloop}[4][]{%
        \draw[#1] let \p1 = ($(#3)-(#2)$) in (#3) arc (#4:({#4+180}):({0.5*veclen(\x1,\y1)});)
}

\newcommand{\valencia}{Instituto de F\'{\i}sica Corpuscular, Universitat de Val\`{e}ncia - 
Consejo Superior de Investigaciones Cient\'{\i}ficas,\\ 
Parc Cient\'{\i}fic, E-46980 Paterna, Valencia, Spain}

\begin{document}

\title{From Jacobi off-shell currents to integral relations}
\author[a]{Jos\'e Llanes Jurado,}
\author[a]{Germ\'an Rodrigo,}
\author[a]{and William J.~Torres~Bobadilla}
\affiliation[a]{\valencia}
\emailAdd{jollaju@alumni.uv.es}
\emailAdd{german.rodrigo@csic.es}
\emailAdd{william.torres@ific.uv.es}
\preprint{IFIC/17-17}
\abstract{
In this paper, we study off-shell currents built from the Jacobi identity of 
the kinematic numerators of $gg\to X$ with $X=ss,q\bar{q},gg$. We find that 
these currents can be schematically written in terms of three-point interaction Feynman rules. 
This representation allows for a straightforward understanding of the 
Colour-Kinematics duality as well as for the construction of the building blocks for the generation
of higher-multiplicity tree-level and multi-loop numerators.
We also provide one-loop integral relations through the Loop-Tree duality formalism
with potential applications and advantages for the computation of relevant physical 
processes at the Large Hadron Collider. 
We illustrate these integral relations with the explicit examples of 
QCD one-loop numerators of $gg\to ss$.
}

\setcounter{page}{1}

\maketitle

\input{intro.tex}
\input{section2.tex}
\input{section3.tex}
\input{section4.tex}

\input{conclusions.tex}
\input{ackno.tex}
\appendix
\input{appFR.tex}
\input{appProj.tex}

\bibliographystyle{JHEP}
\bibliography{refs}

\end{document}

%% file: intro.tex
\section{Introduction}

The Large Hadron Collider (LHC) programme demands us to refine our understanding in the calculation 
of observables in High Energy Physics. The scattering amplitudes are the backbone of theoretical 
predictions. They, besides of having practical applications in particle physics, also have a mathematical
elegance, whose properties have allowed to develop new techniques to perform calculations that a while 
ago were very cumbersome.


Nevertheless, apart of providing theoretical predictions to phenomenological observables
with the use of these amplitudes, we can use their formal properties.
These properties are in principle hidden at the Lagrangian level
but visible in the scattering amplitudes. In particular, QCD scattering amplitudes can be colour decomposed or 
simply split into two pieces, one containing information of the colour structure and other one taking care of kinematic variables~\cite{Cvitanovic:1980bu,Berends:1987cv,Mangano:1987xk,Kosower:1987ic,Bern:1990ux}. 
The latter is often called primitive amplitude. 
An $n$-point colour-dressed tree-level amplitude generates $n!$ primitive amplitudes, although, 
due to cyclic invariance the number of independent primitive amplitudes is reduced to $(n-1)!$. 
In the same manner, with a proper choice of the basis for the colour structure~\cite{DelDuca:1999rs,Maltoni:2002mq}, the 
Kleiss-Kuijf relations~\cite{Kleiss:1988ne} reduce the number of independent primitive amplitudes to $(n-2)!$.

Remarkably, it has been found by Bern, Carrasco and Johansson (BCJ)~\cite{Bern:2008qj,Bern:2010ue} 
that scattering amplitudes
in gauge theories satisfy a colour-kinematic dual representation. 
This representation states that the kinematic part of numerators of 
Feynman diagrams obeys Jacobi identities and anti-symmetric
relations, similar to the corresponding structure constants and group generators 
of the colour algebra, see Fig.~\ref{fig:Jac}.
Therefore, the rearrangement of numerators, according to these identities,
leads to a set of relations between primitive amplitudes, called
BCJ relations, reducing further the number of independent primitive amplitudes to $(n-3)!$.

\begin{figure}[htb]
\centering
\includegraphics[scale=1]{figs/fig1.epsi}
\caption{Colour anti-symmetry relation and Jacobi identity.}
\label{fig:Jac}
 \end{figure}

Several calculations have been done with the use of the colour-kinematics duality (CKD), 
going from supersymmetric theories, up to five-loop level~\cite{Bern:2010ue,Bern:2012uf,Carrasco:2011mn,Bjerrum-Bohr:2013iza,Carrasco:2012ca,Bern:2011rj,BoucherVeronneau:2011qv,Mastrolia:2012wf,Yang:2016ear,Bern:2017ucb}, to 
non-supersymmetric ones, up to two-loop level~\cite{Bern:2013yya,Nohle:2013bfa,Badger:2015lda}. 
Similarly, new relations at one-loop level have been found with a clever use of the BCJ relations
with string theory~\cite{BjerrumBohr:2010zs,Tourkine:2016bak,Ochirov:2017jby,He:2017spx} and unitarity based methods~\cite{Chester:2016ojq,Primo:2016omk}.\\ 

In this paper we follow a diagrammatic approach to construct compact expressions for off-shell currents
built from the Jacobi identity of kinematic numerators. First, we focus on the $2\to2$ processes considered in Ref.~\cite{Mastrolia:2015maa} by one of the current
authors, $gg\to X$ with $X = ss,q\bar{q},gg$, finding that CKD 
can be cast into a systematic representation, 
whose shape relies on three-point interactions. A pictorial representation is given in terms of Feynman rules. 
This compact representation of off-shell currents allows for a straightforward generation of 
CKD tree-level amplitudes, by following the algorithms based on gauge transformations~\cite{Bern:2010yg,BjerrumBohr:2012mg,Boels:2012sy,Mastrolia:2015maa}. 

Later, we investigate the properties of these off-shell currents when they are embedded in a richer topology, namely 
higher-multiplicity or multi-loop scattering amplitudes. 
It turns out that with a proper decomposition of the four off-shell momenta in on-shell massless ones we find a simple structure that allows to write objects built from CKD off-shell currents in terms of at most two squared off-shell momenta. We remark that this decomposition, showing the full off-shell dependence, does not have additional contributions of squared momenta. 

As a byproduct of the off-shell decomposition at one-loop level, we show that new relations
among Feynman integrals with the same number of propagators emerge. 
In order to extract these identities, we use the Loop-Tree duality (LTD) formalism~\cite{Catani:2008xa,Sborlini:2016hat,Sborlini:2016gbr,Hernandez-Pinto:2015ysa,Buchta:2015wna}. 
To this end, we consider, for illustrative reasons, the one-loop process of $gg\to ss$, whose numerators
are built from Jacobi off-shell currents. We note that numerators with higher rank in the loop momentum  are written in terms of lower ones.  

Algebraic manipulations have been carried out by using the \mathematica~packages \Feynarts~\cite{Hahn:2000kx}
and \Feyncalc~\cite{Mertig:1990an, Shtabovenko:2016sxi}.

%% file: section2.tex
\section{Colour-Kinematics duality}
\label{sec:preliminaries}

In this section, we set up the notation and normalisation used throughout
this paper. Let us first consider an $m$-point tree-level amplitude
\begin{align}
\mathcal{A}_{m}^{\text{tree}}\left(1,2,\ldots,m\right) & =\sum_{i=1}^{N}\frac{c_{i}\,n_{i}}{D_{i}}\,\qquad D_{i}=\prod_{\alpha_{i}}s_{\alpha_{i}}\,,
\end{align}
where the sum runs over all diagrams $i$ with only cubic vertices,
$c_{i}$ are the colour factors, $n_{i}$ the kinematic numerators,
and $D_{i}$ collect the denominators of all internal propagators.
As we shall see in the next section, contact terms are absorbed into
cubic diagrams once they are replaced with numerator factors cancelling
propagators, i.e., $s_{\alpha}/s_{\alpha}$ and assigning their contribution
to the proper diagram according to the colour factor. 

The main property of the colour factors is that they satisfy the Jacobi
identities
\begin{subequations}
\begin{align}
-\tilde{f}^{a_{1}a_{2}x}\tilde{f}^{a_{3}a_{4}x}-\tilde{f}^{a_{1}a_{4}x}\tilde{f}^{a_{2}a_{3}x}+\tilde{f}^{a_{1}a_{3}x}\tilde{f}^{a_{2}a_{4}x} & =0\,,\\
-\tilde{f}^{a_{1}a_{2}x}T^{x}-T^{a_{1}}T^{a_{2}}+T^{a_{2}}T^{a_{1}} & =0\,,
\end{align}
\label{eq:JacExpl}
\end{subequations}
where we have adopted the normalisation $\tilde{f}^{abc}=\text{Tr}([T^{a},T^{b}]T^{c})=i\sqrt{2}f^{abc}$
and $T^{a}=\sqrt{2}t^{a}$, with $f^{abc}$ and $t^{a}$ the standard
structure constants and generators of $\mathsf{SU(N)}$, to avoid prefactors in the
next calculations. 

Furthermore, for any $m$-point amplitude, we can always find three
colour factors built from (\ref{eq:JacExpl}), say 
\begin{align}
c_{i}=\ldots\tilde{f}^{a_{1}a_{2}x}\tilde{f}^{a_{3}a_{4}x}\ldots\,, &  & c_{j}=\dots\tilde{f}^{a_{1}a_{4}x}\tilde{f}^{a_{2}a_{3}x}\ldots\,, &  & c_{k}=\ldots\tilde{f}^{a_{1}a_{3}x}\tilde{f}^{a_{2}a_{4}x}\ldots\,,
\end{align}
where the `$\ldots$' state for common terms in the three colour factors.
Therefore, the Jacobi identity takes the form
\begin{align}
-c_{i}-c_{j}+c_{k} & =0\,.\label{eq:jac1}
\end{align}

Since colour factors and kinematic numerators satisfy anti-symmetry
relations under a swapping of legs, $c_{j}\to-c_{j}\Rightarrow n_{j}\to-n_{j}$,
we promote~\eqref{eq:jac1} to be dual in the kinematic sector, 
\begin{align}
 & -n_{i}-n_{j}+n_{k}\,.\label{eq:Joff}
\end{align}
This relation between colour factors and kinematic numerators is referred
to as Colour-Kinematics duality (CKD). CKD is always satisfied
for $2\to2$ processes at tree-level. However, its generalisation
for processes at higher-multiplicity or multi-loop level is not straightforward.
In fact, CKD for tree-level amplitudes can always be recovered
by rearranging numerators but at multi-loop level it has
not been proven yet remaining as a conjecture~\cite{Bern:2010tq}.

\section{Jacobi off-shell currents from the colour-kinematics duality}
\label{sec:offcurrents}

In this section, we compute off-shell currents built from the Jacobi-like combination of
numerators. We recap the diagrammatic approach of Ref.~\cite{Mastrolia:2015maa},
used by one of the current authors. We find that this combination of numerators
for the processes $gg\to X$, with $X=ss, q\bar{q}, gg$ can always be schematically 
represented in terms of three-point interaction Feynman rules.

These off-shell currents are computed in the axial gauge, in which the polarisation tensor
of the gluon propagator is written as 
\begin{align}
\Pi^{\alpha\beta}\left(p_{i},q\right)=-g^{\alpha\beta}+\frac{p_{i}^{\alpha}q^{\beta}+p_{i}^{\beta}q^{\alpha}}{p_{i}\cdot q}\,,
\label{eq:gluonprop}
\end{align}
where $p_i$ is the momentum of the internal (or off-shell) gluon and $q$ its gauge reference momentum, such that 
$q^2=0$ and $p_i\cdot q\ne 0$. Throughout this paper the reference momentum $q$ is chosen to be 
the same for all internal gluons. Although, this choice of gauge does not play any role for $2\to 2$ processes
at tree-level it allows for simplifications when considering processes with higher-multiplicity at tree
or multi-loop level, as we shall observe in Sec.~\ref{sec:decomposition}.

CKD is studied for the processes depicted in Fig.~\ref{fig:JacProcesses}.
We consider quark (anti-quark) living in the fundamental (anti-fundamental) representation
of $\mathsf{SU(N)}$, while scalars and gluons are in the adjoint one. 

\begin{figure}[htb!]
\centering
\includegraphics[scale=1]{figs/fig2.epsi}
\caption{Jacobi combination for $gg\to X$ with $X=ss,q\bar{q},gg$.}
\label{fig:JacProcesses}
\end{figure}

In Fig.~\ref{fig:JacProcesses}, $n(\hdots)$ is understood as the numerator of the Feynman diagram,
in which scalar propagators, $1/p_i^2$, have been removed. 
Nevertheless, contributions of $1/(p_i\cdot q)$ are retained in order to work out with same dimensions.
The four-point interactions, or simply contact terms, 
$ssgg$ and $gggg$ have been rewritten in terms 
of cubic diagrams only. Thus, a numerator containing one contact term can be expressed as
\begin{align}
n_4 = s_{\alpha_i}n_{4:i}c_i + s_{\alpha_j}n_{4:j}c_j + s_{\alpha_k}n_{4:k}c_k\,.
\end{align} 

We compute the Jacobi off-shell currents for the processes $gg\to ss, gg\to q\bar{q}$
and $gg\to gg$ in axial gauge. In order to elucidate the dependence on the gauge, 
we split the propagator~\eqref{eq:gluonprop} into Feynman and covariant parts, i.e.,
\begin{align}
\Pi^{\mu\nu}(p_i,q) = \Pi_{\text{Fey}}^{\mu\nu} + \Pi_{\text{Ax}}^{\mu\nu}(p_i,q)\,,
\end{align}
with $\Pi_{\text{Fey}}^{\mu\nu} = -g^{\mu\nu}$ and 
$\Pi_{\text{Ax}}^{\mu\nu}(p_i,q) = \frac{p_i^\mu q^\nu+p_i^\nu q^\mu}{p_i\cdot q}$.
This splitting allows us to write the Jacobi off-shell currents as
\begin{align}
J_{\text{x}}^{\mu_{1}\hdots\mu_{4}} = J_{\text{x-Fey}}^{\mu_{1}\hdots\mu_{4}} + J_{\text{x-Ax}}^{\mu_{1}\hdots\mu_{4}}\,,
\qquad\text{with x$=$s,q,g}\,.
\end{align}

In the following, we list the Jacobi off-shell currents.
\subsection{$gg\to ss$}

We start considering the process $gg\to ss$ with gluons and scalars in the adjoint representation.
The Jacobi identity for kinematic numerator takes the form,
\begin{align}
J_{\text{s-Fey}}^{\mu_{1}\mu_{4}} =  & p_{1}^{\mu_{1}}
\Bigg(\parbox{20mm}{\input{figs/GSS4.tex}}\Bigg) - (1\leftrightarrow4)\,,  \notag\\[-0.5cm]
J_{\text{s-Ax}}^{\mu_{1}\mu_{4}} =  &
\frac{1}{q\cdot p_{14}}\left[\mathcal{P}_{1}^{\mu_{1}\mu_{4}}-(1\leftrightarrow4)\right]q^\sigma\Bigg(\parbox{20mm}{\input{figs/GSS23q.tex}} \Bigg) 
 +\frac{1}{q\cdot p_{23}} \left(p_{2}^{2}- p_{3}^{2}\right)
 q^\sigma\Bigg(\parbox{20mm}{\input{figs/GGG41.tex}}\Bigg)\,,
\label{eq:JsOff}
\end{align}
where $p_{ij}^\alpha\equiv\left(p_i+p_j\right)^\alpha$ and 
$\mathcal{P}_i^{\mu_{i}\mu_{j}}\equiv p_{i}^{\mu_{i}}p_{i}^{\mu_{j}}-p_{i}^{2}\,g^{\mu_{i}\mu_{j}}$.
The Feynman rules for the three-point interaction vertices are defined in Appendix~\ref{app:FRs}.

We stress that the Feynman rules in Eq.~\eqref{eq:JsOff} are depicted to provide a simple
and compact representation of the Jacobi off-shell current. 
In particular, contributions to $J_{\text{s-Fey}}$ do not obey momentum
conservation. It only follows the structure of the Feynman rule for $gss$, Eq.~\eqref{eq:gss}. 
On the other hand, contributions to $J_{\text{s-Ax}}$ do obey momentum conservation. 

We remark that contributions to $J_{\text{s-Fey}}$ come from the three diagrams of Fig.~\ref{fig:JacProcesses}.a, 
whereas $J_{\text{s-Ax}}$ gets only one contribution from the diagram with an 
internal gluon. Additionally, we note that CKD for the former is satisfied when the 
external gluons are set on-shell, or simply ask for transversality condition, $\varepsilon(p_i)\cdot p_i=0$.
In the latter, both gluons and scalars have to be set on-shell, $p_i = 0$.

\subsection{$gg \to q\bar{q}$}

We now consider the process $gg \to q\bar{q}$ with quark (anti-quark) in the fundamental representation.
The Jacobi off-shell current $J_{\text{q}}$ takes the form, 
\begin{align}
J_{\text{q-Fey}}^{\mu_{1}\mu_{4}}& = \Bigg[p_{1}^{\mu_{1}}
\Bigg(\parbox{20mm}{\input{figs/FFG4.tex}}\Bigg)
- (1\leftrightarrow4)\Bigg]
+\Bigg[\slashed p_{2}\Bigg(\parbox{20mm}{\input{figs/FGG3.tex}}\Bigg)
+ (2\leftrightarrow3)\Bigg]
\notag\\[-0.5cm]
J_{\text{q-Ax}}^{\mu_{1}\mu_{4}}& =\frac{1}{q\cdot p_{14}}\Bigg[\mathcal{P}_{1}^{\mu_{1}\mu_{4}}-(1\leftrightarrow4)\Bigg]q^\sigma\Bigg(\parbox{20mm}{\input{figs/FFG23q.tex}}\Bigg) 
+\frac{\slashed p_{23}}{q\cdot p_{23}} 
q^\sigma\Bigg(\parbox{20mm}{\input{figs/GGG41.tex}} \Bigg)\,.
\label{eq:JqOff}
\end{align}
In Eq.~\eqref{eq:JqOff} we have included an additional interaction between one quark and two gluons,~\eqref{eq:qgg}. This Feynman rule does not obey momentum conservation and does not have any 
physical meaning. It nevertheless allows for a universality in the decomposition of the off-shell current 
generated from CKD. 

Similar to $J_{\text{s}}$, contributions to $J_{\text{q-Fey}}$ come from the three diagrams of 
Fig.~\ref{fig:JacProcesses}.b, whereas, $J_{\text{q-Ax}}$ gets contribution from the diagram with an
internal gluon. We also see that in order to recover CKD the four particles have to be set on-shell.
For this case, apart from asking for transversality conditions of external gluons we also need 
external quark (anti-quark) to satisfy Dirac equations, $\slashed p_2\bar{u}(p_2)=v(p_3)\slashed p_3=0$.\\

The exchanging $(2\leftrightarrow3)$ in~Eq.\eqref{eq:JqOff} takes also care of the Dirac indices. 
Hence, any expression containing gamma matrices should intrinsically be read as follows,
\begin{align}
\slashed p_{2}\left[\gamma^{\mu_{1}},\gamma^{\mu_{4}}\right]\equiv(\slashed p_{2})_{i_{2} \bar{k}}[\gamma^{\mu_{1}},\gamma^{\mu_{4}}]_{k\,\bar{\j}_{3}}\,,
\end{align}
therefore, 
\begin{align}
\slashed p_{2}\left[\gamma^{\mu_{1}},\gamma^{\mu_{4}}\right]\underset{2\leftrightarrow3}{\longrightarrow}\left[\gamma^{\mu_{4}},\gamma^{\mu_{1}}\right]\slashed p_{3}\,.
\end{align}

\subsection{$gg\to gg$}

As last application, we consider the process of $gg\to gg$. 
Since there is no mixing between different kind of particles, we end up with the most 
compact and symmetric expression for the Jacobi off-shell current $J_{\text{g}}$, 
\begin{align}
J_{\text{g-Fey}}^{\mu_{1}\mu_{2}\mu_{3}\mu_{4}} & =\sum_{\sigma\in \mathsf{Z_{4}}}
p_{\sigma_1}^{\mu_{\sigma_1}} \Bigg(\;\parbox{20mm}{\input{figs/GGG.generic.tex}}\;\;\Bigg)\,,
\notag\\[-0.5cm]
J_{\text{g-Ax}}^{\mu_{1}\mu_{2}\mu_{3}\mu_{4}} & = \sum_{\sigma\in\mathsf{ A_{4}}}
\frac{\mathcal{P}_{\sigma_{1}}^{\mu_{\sigma_{1}}\mu_{\sigma_{2}}}}{q\cdot p_{\sigma_{1}\sigma_{2}}}
q^{\alpha}\Bigg(\;\parbox{20mm}{\input{figs/GGG.genericq.tex}}\;\;\Bigg)\,,
\label{eq:JgOff}
\end{align}
where the sums run over the all the permutations of the cyclic ($\mathsf{Z_4}$) 
and alternative ($\mathsf{A_4}$)
groups, respectively, whose elements are represented by $\sigma_i$. 

As in the previous cases, $J_{\text{g-Fey}}$ comes from the three diagrams of Fig.~\ref{fig:JacProcesses}.c
and their contributions in terms of Feynman rules do not  
obey momentum conservation as they do for the ones in $J_{\text{g-Ax}}$.
Furthermore, we observe a similar pattern between $J_{s}$ and $J_{g}$. In fact, the former can be 
easily recovered from the latter by setting $p_i^{\mu_2}=p_i^{\mu_3}=0$ and $g^{\mu_2\mu_3}=1$, 
for $i=1,\hdots,4$.\\

From $J_{\text{s,q,g}}$, Eq.s~(\ref{eq:JsOff},\ref{eq:JqOff}) and~\eqref{eq:JgOff}, 
we observe that Jacobi off-shell currents can 
systematically be cast into objects whose structure follow the 
three-point interaction Feynman rules. We remark that all $J_{\text{x-Fey}}$ are obtained from the Jacobi identity 
of three kinematic numerators of Eq.~\eqref{eq:Joff} 
and their structure, after algebraic manipulations, is always written
as Feynman rules where momentum conservation is not preserved. 
Likewise, CKD is straightforwardly  recovered when the four particles in the off-shell currents are set on-shell. 
On the other hand, the Feynman rules appearing in all $J_{\text{x-Ax}}$ do obey momentum
conservation and the way how CKD is satisfied is individually at the level of diagrams. 
It is indeed for this reason that the single diagram with an internal gluon for $gg\to ss$ and 
$gg\to q\bar{q}$ vanishes by its own when the four particles are set on-shell. 
In fact, contributions from the covariant part of the polarisation tensor of the gluon propagator~\eqref{eq:gluonprop}, 
$q^\mu p^\nu + q^\nu p^\mu$, manifest the same pattern, 
\begin{align}
p_{\sigma_{1}}^{\mu_{\sigma_{1}}}p_{\sigma_{1}}^{\mu_{\sigma_{2}}}-p_{\sigma_{1}}^{2}g^{\mu_{\sigma_{1}}\mu_{\sigma_{2}}}\,,
\end{align}
which vanishes when it is contracted with $\varepsilon^{\mu_{\sigma_{i}}}(p_{\sigma_{i}})$ and on-shellness of  $p_{\sigma_{i}}$ is imposed, $p_{\sigma_{i}}^2=0$.


%% file: figs/GSS4.tex
\unitlength=0.20bp%
\begin{feynartspicture}(300,300)(1,1)
\FADiagram{}
\FAProp(3.,10.)(10.,10.)(0.,){/Cycles}{0}
\FAProp(16.,15.)(10.,10.)(0.,){/ScalarDash}{0}
\FAProp(16.,5.)(10.,10.)(0.,){/ScalarDash}{0}
\FAVert(10.,10.){0}
\FALabel(16.2273,15.5749)[cb]{\tiny $p_2$}
\FALabel(16.2273,4.5749)[ct]{\tiny $p_3$}
\FALabel(2.3,8.93)[ct]{\tiny $p_4, \mu_4$}
\end{feynartspicture}

%% file: figs/GSS23q.tex
\unitlength=0.20bp%
\begin{feynartspicture}(300,300)(1,1)
\FADiagram{}
\FAProp(3.,10.)(10.,10.)(0.,){/Cycles}{0}
\FAProp(16.,15.)(10.,10.)(0.,){/ScalarDash}{0}
\FAProp(16.,5.)(10.,10.)(0.,){/ScalarDash}{0}
\FAVert(10.,10.){0}
\FALabel(16.2273,15.5749)[cb]{\tiny $p_2$}
\FALabel(16.2273,4.5749)[ct]{\tiny $p_3$}
\FALabel(2.3,8.93)[ct]{\tiny $-p_{23}, \sigma$}
\end{feynartspicture}

%% file: figs/GGG41.tex
\unitlength=0.20bp%
\begin{feynartspicture}(300,300)(1,1)
\FADiagram{}
\FAProp(3.,10.)(10.,10.)(0.,){/Cycles}{0}
\FAProp(16.,15.)(10.,10.)(0.,){/Cycles}{0}
\FAProp(16.,5.)(10.,10.)(0.,){/Cycles}{0}
\FALabel(16.2273,15.5749)[cb]{\tiny $p_4, \mu_4$}
\FALabel(16.2273,4.5749)[ct]{\tiny $p_1, \mu_1$}
\FALabel(2.3,8.93)[ct]{\tiny $-p_{14}, \sigma$}
\FAVert(10.,10.){0}
\end{feynartspicture}

%% file: figs/FFG4.tex
\unitlength=0.20bp%
\begin{feynartspicture}(300,300)(1,1)
\FADiagram{}
\FAProp(3.,10.)(10.,10.)(0.,){/Cycles}{0}
\FAProp(16.,15.)(10.,10.)(0.,){/Straight}{-1}
\FAProp(16.,5.)(10.,10.)(0.,){/Straight}{1}
\FAVert(10.,10.){0}
\FALabel(16.2273,15.5749)[cb]{\tiny $p_2$}
\FALabel(16.2273,4.5749)[ct]{\tiny $p_3$}
\FALabel(2.3,8.93)[ct]{\tiny $p_4, \mu_4$}
\end{feynartspicture}

%% file: figs/FGG3.tex
\unitlength=0.20bp%
\begin{feynartspicture}(300,300)(1,1)
\FADiagram{}
\FAProp(3.,10.)(10.,10.)(0.,){/Straight}{1}
\FAProp(16.,15.)(10.,10.)(0.,){/Cycles}{0}
\FAProp(16.,5.)(10.,10.)(0.,){/Cycles}{0}
\FAVert(10.,10.){0}
\FALabel(16.2273,15.5749)[cb]{\tiny $p_4, \mu_4$}
\FALabel(16.2273,4.5749)[ct]{\tiny $p_1, \mu_1$}
\FALabel(2.3,8.93)[ct]{\tiny $p_3$}
\end{feynartspicture}

%% file: figs/FFG23q.tex
\unitlength=0.20bp%
\begin{feynartspicture}(300,300)(1,1)
\FADiagram{}
\FAProp(3.,10.)(10.,10.)(0.,){/Cycles}{0}
\FAProp(16.,15.)(10.,10.)(0.,){/Straight}{-1}
\FAProp(16.,5.)(10.,10.)(0.,){/Straight}{1}
\FAVert(10.,10.){0}
\FALabel(16.2273,15.5749)[cb]{\tiny $p_2$}
\FALabel(16.2273,4.5749)[ct]{\tiny $p_3$}
\FALabel(2.3,8.93)[ct]{\tiny $-p_{23}, \sigma$}
\end{feynartspicture}

%% file: figs/GGG.generic.tex
\unitlength=0.20bp%
\begin{feynartspicture}(300,300)(1,1)
\FADiagram{}
\FAProp(3.,10.)(10.,10.)(0.,){/Cycles}{0}
\FAProp(16.,15.)(10.,10.)(0.,){/Cycles}{0}
\FAProp(16.,5.)(10.,10.)(0.,){/Cycles}{0}
\FALabel(16.2273,15.5749)[cb]{\tiny $p_{\sigma_2}, \mu_{\sigma_2}$}
\FALabel(16.2273,4.5749)[ct]{\tiny $p_{\sigma_3}, \mu_{\sigma_3}$}
\FALabel(2.3,8.93)[ct]{\tiny $p_{\sigma_4}, \mu_{\sigma_4}$}
\FAVert(10.,10.){0}
\end{feynartspicture}

%% file: figs/GGG.genericq.tex
\unitlength=0.20bp%
\begin{feynartspicture}(300,300)(1,1)
\FADiagram{}
\FAProp(3.,10.)(10.,10.)(0.,){/Cycles}{0}
\FAProp(16.,15.)(10.,10.)(0.,){/Cycles}{0}
\FAProp(16.,5.)(10.,10.)(0.,){/Cycles}{0}
\FALabel(16.2273,15.5749)[cb]{\tiny $p_{\sigma_3}, \mu_{\sigma_3}$}
\FALabel(16.2273,4.5749)[ct]{\tiny $p_{\sigma_4}, \mu_{\sigma_4}$}
\FALabel(2.3,8.93)[ct]{\tiny $-p_{\sigma_3\sigma_4}, \alpha$}
\FAVert(10.,10.){0}
\end{feynartspicture}

%% file: section3.tex
\section{Colour-Kinematics duality for multi-leg amplitudes}
\label{sec:decomposition}

In Sec.~\ref{sec:offcurrents} we constructed off-shell currents from the Jacobi identity of numerators.
Hence, in order to understand their behaviour when they are embedded in a richer topology, 
trees with higher-multiplicity or  multi-loop level, we plug external wave functions
that do not necessary have to be on-shell. 
We find that numerators built from Jacobi off-shell currents, $J_{\text{s,q,g}}$, are always written in terms 
of at most the product of two squared momenta. Being these momenta the ones attached to the off-shell
current.

\subsection{Momentum decomposition in terms of on-shell momenta}

Let us decompose an off-shell momentum $p_i$ into two on-shell momenta,
\begin{align}
p_{i}^{\alpha}=r_{i}^{\alpha}+\frac{p_{i}^{2}}{2q\cdot r_{i}}q^{\alpha}\,,
\label{eq:offshell}
\end{align}
where $r_i$ and $q$ are massless momenta, $r_i^2=q^2=0$, with the condition
 $r_i\cdot q \ne 0 $.  For the purpose of our calculations, $q$ is chosen to be same as the reference
 momenta of the gluon propagator in the axial gauge~\eqref{eq:gluonprop}.
  
 Hence, completeness relations for polarisation vectors, in axial gauge, and 
 spinors take the form
 \begin{align}
 \sum_{\lambda=1}^{d_{s}-2}\varepsilon_{\lambda\left(d_{s}\right)}^{\alpha}\left(p_{i}\right)\varepsilon_{\lambda\left(d_{s}\right)}^{*\beta}\left(p_{i}\right)&=-g^{\alpha\beta}+\frac{r_{i}^{\alpha}q^{\beta}+r_{i}^{\beta}q^{\alpha}}{r_{i}\cdot q}+\frac{p_{i}^{2}}{\left(r_{i}\cdot q\right)^{2}}q^{\alpha}q^{\beta}\,,
 \label{eq:gluonCR}\\
\sum_{\lambda=1}^{2^{(d_{s}-2)/2}}u_{\lambda\left(d_{s}\right)}\left(p_{i}\right)\bar{u}_{\lambda\left(d_{s}\right)}\left(p_{i}\right)&=\slashed r_{i}+\frac{p_{i}^{2}}{2r_{i}\cdot q}\slashed q\,.\label{eq:fermionCR}
 \end{align}

The completeness relations~\eqref{eq:gluonCR} and~\eqref{eq:fermionCR} distinguish between
on- and off-shell quantities. The on-shell ones account for the numerator of  gluon and fermion 
propagators, whereas the off-shell ones take care of contributions coming from $p_i^2$ only. 
Therefore, these completeness relations can be cast into, 
\begin{subequations}
\begin{align}
\sum_{\lambda=1}^{d_{s}-2}\varepsilon_{\lambda\left(d_{s}\right)}^{\alpha}\left(p_{i}\right)\varepsilon_{\lambda\left(d_{s}\right)}^{*\beta}\left(p_{i}\right)&=\sum_{\lambda_{i}=1}^{d_{s}-2}\varepsilon_{i}^{\alpha}\varepsilon_{i}^{*\beta}+\frac{p_{i}^{2}}{\left(r_{i}\cdot q\right)^{2}}q^{\alpha}q^{\beta}\,,\\\sum_{\lambda=1}^{2^{(d_{s}-2)/2}}u_{\lambda\left(d_{s}\right)}\left(p_{i}\right)\bar{u}_{\lambda\left(d_{s}\right)}\left(p_{i}\right)&=\sum_{\lambda_{i}=1}^{2^{(d_{s}-2)/2}}u_{i}\bar{u}_{i}+\frac{p_{i}^{2}}{2r_{i}\cdot q}\slashed q\,.
\end{align}
\label{eq:offnum}
\end{subequations}
Here and in the following we use the abbreviations
\begin{align}
&\varepsilon_{i}^{\mu}\equiv\varepsilon_{\lambda_{i}\left(d_{s}\right)}^{\mu}\left(r_{i}\right)\,,
&&\bar{u}_{i}\equiv\bar{u}_{\lambda_{i}\left(d_{s}\right)}\left(r_{i}\right)\,,
&&v_{i}\equiv v_{\lambda_{i}\left(d_{s}\right)}\left(r_{i}\right)\,.
\end{align}

As mentioned above, we have chosen $q$ to be the same for all off-shell momenta. 
This is done to remove, as much as possible, redundant terms in the following calculations. 
Thus, besides the on-shell conditions $r_i$ satisfy, $\varepsilon_i\cdot r_i=0$ and 
$\bar{u}_i\slashed r_i = \slashed r_i v_i=0$, we also have, 
as a consequence of decomposition~\eqref{eq:offshell},  $\varepsilon_i\cdot q=\varepsilon_i\cdot p_i=0$. 

\subsection{Construction of numerators from Jacobi off-shell currents}
\label{sec:embedding}

In order to study the behaviour of numerators built from the Jacobi off-shell currents, we consider
\begin{subequations}
\begin{align}
&N_{\text{s}}=N_{\text{s}\,\mu_{1}\mu_{4}}X^{\mu_{1}\mu_{4}}\,,&&N_{\text{s}\,\mu_{1}\mu_{4}}=J_{\text{s}}^{\nu_{1}\nu_{4}}\Pi_{\mu_{1}\nu_{1}}\left(p_{1},q\right)\Pi_{\mu_{4}\nu_{4}}\left(p_{4},q\right)\,,\\
&N_{\text{q}}=N_{\text{q}\,\mu_{1}\mu_{4}}X^{\mu_{1}\mu_{4}}\,,&&N_{\text{q}\,\mu_{1}\mu_{4}}=\slashed p_{2}J_{\text{q}}^{\nu_{1}\nu_{4}}\slashed p_{3}\Pi_{\mu_{1}\nu_{1}}\left(p_{1},q\right)\Pi_{\mu_{4}\nu_{4}}\left(p_{4},q\right)\,,
\label{eq:NumoffshellB}\\
&N_{\text{g}}=N_{\text{g}\,\mu_{1}\hdots\mu_{4}}X^{\mu_{1}\hdots\mu_{4}}\,,&&N_{\text{g}\,\mu_{1}\hdots\mu_{4}}=J_{\text{g}}^{\nu_{1}\hdots\nu_{4}}\Pi_{\mu_{1}\nu_{1}}\left(p_{1},q\right)\hdots\Pi_{\mu_{4}\nu_{4}}\left(p_{4},q\right)\,.
\end{align}
\label{eq:Numoffshell}
\end{subequations}
The tensors $X$ carry the information related to the kinematic part where the off-shell 
currents $J_{\text{s,q,g}}$ are embedded.
They can generate either tree-level or multi-loop numerators.

In the case that any of the gluons attached to the Jacobi off-shell current is set on-shell, 
the polarisation tensor of the gluon should be substituted by the corresponding 
polarisation vector. Analogously for quarks, the $\slashed{p}_i$ in Eq.~\eqref{eq:NumoffshellB} should be substituted 
by the corresponding spinor. 

Let us begin by defining a shorthand notation for the definition of the numerators
\begin{align}
&\mathcal{E}_{ij}^{\nu_{i}\nu_{j}}\equiv\sum_{\lambda_{i},\lambda_{j}=1}^{d_{s}-2}\varepsilon_{i}^{\nu_{i}}\varepsilon_{j}^{\nu_{j}}\,,&&\mathcal{Q}_{i}^{\nu_{i}\nu_{j}}\equiv\sum_{\lambda_{j}=1}^{d_{s}-2}q^{\nu_{i}}\varepsilon^{\nu_{j}}\,,&&\mathfrak{q}^{\nu_{i}\nu_{j}}\equiv q^{\nu_{i}}q^{\nu_{j}}\,,&&\slashed r_{i}\equiv\sum_{\lambda_{i}=1}^{2^{(d_{s}-2)/2}}u_{i}\bar{u}_{i}\,.
\end{align}
In the following, we list the numerators~\eqref{eq:Numoffshell} built from the Jacobi off-shell currents.

\subsubsection{$gg\to ss$}
We start considering the generic structure of numerators built from the Jacobi off-shell current
$J_{s}$, 
\begin{align}
N_{\text{s}}^{\mu_{1}\mu_{4}} &=\sum_{i,j=1,4}\epsilon_{ij}\,p_{i}^{2}\left(A_{ij}\,\mathcal{E}_{ij}^{\mu_{i}\mu_{j}} +B_{ij}\,\mathcal{Q}_{i}^{\mu_{i}\mu_{j}}+C_{ij}\,p_{j}^{2}\,\mathfrak{q}^{\mu_{i}\mu_{j}}\right)+\left(p_{2}^{2}-p_{3}^{2}\right)\tilde{A}_{14}\,\mathcal{E}_{14}^{\mu_{1}\mu_{4}}\,,
\label{eq:JsOffDec}
\end{align}
with $\epsilon_{ij}$ the Levi-Civita tensor with the signature $\epsilon_{14} = +1$ and 
\begin{align}
&&A_{ij}=\frac{q\cdot(r_{2}-r_{3})}{q\cdot r_{23}}\varepsilon_{i}\cdot\varepsilon_{j}\,,
&&&B_{ij}=-\frac{1}{q\cdot r_{i}}\left[\frac{q\cdot(r_{2}-r_{3})}{q\cdot r_{23}}\varepsilon_{j}\cdot r_{i}+\varepsilon_{j}\cdot(r_{2}-r_{3})\right]\,,\nn
&&\tilde{A}_{14}=\frac{q\cdot(r_{1}-r_{4})}{q\cdot r_{14}}\varepsilon_{1}\cdot\varepsilon_{4}\,,&&&C_{ij}=\frac{q\cdot(r_{2}-r_{3})q\cdot(r_{i}-r_{j})}{\left(q\cdot r_{1}\right){}^{2}\left(q\cdot r_{4}\right){}^{2}}\,.
\end{align}
From Eq.~\eqref{eq:JsOffDec}, we do not get terms proportional to $p_i^2p_j^2$ with $i=1,\hdots,4, j=2,3$ and $i\ne j$, because $q\cdot(p_4-p_1) = q\cdot(r_4-r_1)$.

\subsubsection{$gg \to q\bar{q}$}
We now turn our attention to numerators built from the Jacobi off-shell current $J_{q}$, 
\begin{align}
N_{\text{q}}^{\nu_{1}\nu_{4}}=&\Bigg\{p_{1}^{2}\Bigg[\frac{\slashed r_{2}\slashed q\slashed r_{3}}{q\cdot r_{23}}\left(\varepsilon_{1}\cdot\varepsilon_{4}\mathcal{E}_{14}^{\nu_{1}\nu_{4}}-\frac{\varepsilon_{4}\cdot r_{1}}{q\cdot r_{1}}\mathcal{Q}_{4}^{\nu_{1}\nu_{4}}+p_{4}^{2}\frac{q\cdot r_{23}\,q\cdot(r_{1}-r_{4})}{(q\cdot r_{1}){}^{2}(q\cdot r_{4}){}^{2}}\mathfrak{q}^{\nu_{1}\nu_{4}}\right)\notag\\
&+\frac{1}{q\cdot r_{1}}\left(\slashed r_{2}\slashed\varepsilon_{4}\slashed r_{3}+p_{2}^{2}\frac{\slashed q\slashed\varepsilon_{4}\slashed r_{3}}{2q\cdot r_{2}}+p_{3}^{2}\frac{\slashed r_{2}\slashed\varepsilon_{4}\slashed q}{2q\cdot r_{3}}\right)\mathcal{Q}_{4}^{\nu_{1}\nu_{4}}\Bigg]-\left( 1\leftrightarrow4\right)\Bigg\}\notag\\
&+\Bigg\{\frac{p_{2}^{2}}{2}\Bigg[\slashed\varepsilon_{1}\slashed\varepsilon_{4}\slashed p_{3}\mathcal{E}_{14}^{\nu_{1}\nu_{4}}+\left[\slashed\varepsilon_{1},\slashed q\right]\frac{\slashed r_{3}}{q\cdot r_{4}}p_{4}^{2}\mathcal{Q}_{1}^{\nu_{4}\nu_{1}}\notag\\
&-2\varepsilon_{1}\cdot\varepsilon_{4}q\cdot r_{1}\left(\frac{\slashed r_{3}}{q\cdot r_{23}}+p_{3}^{2}\frac{\slashed q}{4\left(q\cdot r_{2}\right)\left(q\cdot r_{3}\right)}\right)\mathcal{E}_{14}^{\nu_{1}\nu_{4}}-\left(1\leftrightarrow4\right)\Bigg]+\left(2\leftrightarrow3\right)\Bigg\}\,.
\label{eq:JqOffDec}
\end{align}
Unlike the case of the $N_{\text{s}}$, we do not get a simple structure. 
This behaviour is because we are basically mixing two kind of particles,
gluons in the adjoint and quarks in the fundamental representation. The former are governed by 
Lorentz indices whereas the latter accounts for Dirac ones. 
However, we see that numerators $N_{\text{q}}$ can generically be decomposed as
\begin{align}
N_{\text{q}}&=\sum_{i=1}^{4}c_{i}\,p_{i}^{2}+\sum_{\substack{i,j=1\\
i\neq j
}
}^{4}c_{ij}\,p_{i}^{2}p_{j}^{2}\,,
\label{eq:decoGen}
\end{align}
with $c$, coefficients depending on the kinematics where the off-shell current $J_{q}$ is 
embedded in. On top of it, we note a difference with numerators $N_{\text{s}}$, since 
now all powers of $p_i^2p_j^2$, for $i,j=1,\hdots,4$ with $i\ne j$, are present.

\subsubsection{$gg \to gg$}
As last application, we now consider numerators built from the Jacobi off-shell current $J_{g}$.
Since we are working with particles of the same kind, the numerator $N_{\text{g}}$ is
compactly expressed as,
\begin{align}
N_{\text{g}}^{\nu_{1}\hdots\nu_{4}}& = \frac{1}{2}\sum_{i,j,k,l=1}^{4}\epsilon_{ijkl}\,p_{i}^{2}\left(A_{ijkl}\,\mathcal{E}_{ij}^{\nu_{i}\nu_{j}}\mathcal{E}_{kl}^{\nu_{k}\nu_{l}}
+B_{ijkl}\,\mathcal{E}_{jk}^{\nu_{j}\nu_{k}}\mathcal{Q}_{l}^{\nu_{i}\nu_{l}}
+C_{ijkl}\,p_{j}^{2}\,\mathfrak{q}^{\nu_{i}\nu_{j}}\mathcal{E}_{kl}^{\nu_{k}\nu_{l}}\right)\,,
\label{eq:JgOffDec}
\end{align}
with $\epsilon_{ijkl}$ the Levi-Civita tensor with the signature $\epsilon_{1234} = +1$ and 
\begin{align}
A_{ijkl}&=\frac{q\cdot(r_{k}-r_{l})}{q\cdot(r_{k}+r_{l})}\varepsilon_{i}\cdot\varepsilon_{j}\varepsilon_{k}\cdot\varepsilon_{l}\,,\nn
B_{ijkl}&=-\frac{\varepsilon_{j}\cdot\varepsilon_{k}}{q\cdot r_{i}}\left[\frac{q\cdot(r_{j}-r_{k})}{q\cdot(r_{j}+r_{k})}\varepsilon_{l}\cdot r_{i}+\varepsilon_{l}\cdot(r_{j}-r_{k})\right]\,,\nn
C_{ijkl}&=\frac{q\cdot(r_{i}-r_{j})q\cdot(r_{k}-r_{l})}{\left(q\cdot r_{i}\right){}^{2}\left(q\cdot r_{j}\right){}^{2}}\varepsilon_{k}\cdot\varepsilon_{l}\,.
\end{align}
We also observe that numerators built from $J_{g}$ contain all possible combinations of $p_i^2p_j^2$,
similarly to $N_{\text{q}}$ in Eq.~\eqref{eq:decoGen}. \\

We notice from the above results that, because of the way the off-shell wave-functions are parametrised,
there are no terms proportional to $p_i^2p_j^2p_k^2p_l^2$ or $p_i^2p_j^2p_k^2$. 
This is indeed due to the choice of a unique reference momentum $q$ in the definition of 
internal propagators and wave-functions. Therefore, we can claim that any numerator built from the off-shell currents 
$J_{\text{s,q,g}}$ is written in terms of at most 
the product of two squared momenta, $p_i^2p_j^2$. 
Although this observation was done in~\cite{Mastrolia:2015maa}, it is not completely true
for the general case where all reference momenta for the internal gluons are chosen to be 
different. In fact, after the off-shell momenta are decomposed according
to Eq.~\eqref{eq:offshell}, with $q=q_i$, contributions of the product of three and four squared momenta
arise. In particular, scalar products  $\varepsilon_i\cdot q_j$ for $i\ne j$ are 
non-vanishing anymore. 

There are particular cases where CKD can be immediately satisfied
for multi-leg processes. This is the case of the three-gluon vertex
contributions to $X^{\mu_{1}\hdots\mu_{4}}$, in the Feynman gauge.
In fact, if the fouth leg, a gluon, of any Jacobi off-shell current
is set off-shell, we can end up with 
\begin{align}
J_{\text{x-Fey}}^{\mu_{1}\hdots\nu_{45}}\left(p_{1},p_{2},p_{3},p_{45}\right)V_{ggg \nu_{45}}^{\,\mu_{4}\mu_{5}}\left(-p_{45},p_{4},p_{5}\right)\varepsilon_{\mu_{4}}\left(p_{4}\right)\varepsilon_{\mu_{5}}\left(p_{5}\right) & =0\,,\\
J_{\text{x-Fey}}^{\mu_{1}\hdots\nu_{45}}\left(p_{1},p_{2},p_{3},p_{45}\right)\bar{u}\left(p_{4}\right)V_{gq\bar{q} \nu_{45}}\left(-p_{45},p_{4},p_{5}\right)u\left(p_{5}\right) & =0\,,
\end{align}
with $V_{ggg}$ and $V_{gq\bar{q}}$ the standard three-gluon and
quark-anti-quark-gluon vertices.  

It turns out that for $J_{\text{x-Fey}}^{\mu_{1}\hdots\nu_{45}}$,
with x=s,g, we can generate (up to) eight-point numerators built from
these of-shell currents. Whereas, for $J_{\text{q-Fey}}^{\mu_{1}\hdots\nu_{45}}$
we can do the same for (up to) six-point numerator. We do not recover
this kind of identity for the emission of a gluon from a fermion line,
since we get, as expected from the above discussion, a term proportional
to squared momenta.

We stress that $X$'s made of three-gluon vertices miss
all the contributions from contact terms. Nevertheless, adding these contact terms spoils 
the generation of a numerator satisfying CKD. This is due to the fact that diagrams containing 
a contact term have to be slightly modified in order to be consequent with the definition of three-point 
vertices. Indeed, those diagrams have one propagator less and are naively restored by 
inserting factors of $p_i^2/p_i^2$. The proper insertion of contact terms allows for a generation of
numerators satisfying CKD from construction, as was done in Ref.~\cite{Tolotti:2013caa} 
for Yang-Mills theories.

In the same manner, when considering theories with particles living in different representations, 
like gluons and quarks in QCD, it has been shown that the arbitrary procedure of rearranging diagrams
with contact terms does not allow a systematic generation of numerators satisfying CKD. 
In particular, Ref.s~\cite{Melia:2013bta,Melia:2013epa,Johansson:2014zca,Johansson:2015oia} 
discuss the automatic generation
of dual numerators. This relation is straightforwardly satisfied when in the generation of 
the kinematic numerators there are true three-point vertices, i.e.
no contact terms. The latter relies on the emission of single gluons. 
In order to compute processes with mostly quarks one could take advantage of Melia's formalism~\cite{Melia:2013bta}.
This formalism allows us to compute processes with quarks of different flavours and generalise this result
to the case of a single one.

In the next section, we discuss the generation of  one-loop numerators from the Jacobi off-shell currents
of Sec.~\ref{sec:offcurrents}. 
In order to do so, we rely on the Loop-Tree duality formalism.

%% file: section4.tex
\section{Colour-Kinematics duality for $2\to2$ processes at one-loop}

In this section, we study the one-loop $2\to2$ processes when a numerator is built from the Jacobi off-shell
currents~(\ref{eq:JsOff},\ref{eq:JqOff}) and~\eqref{eq:JgOff}. We discuss that at integral level, because of
the decomposition shown in Sec.~\ref{sec:decomposition}, a new set of relations is generated. 
In order to see the structure of these relations, we elaborate on the one-loop process of
$gg\to ss$. 

\subsection{Integral relations}

At one-loop level we have to deal with objects of this form, 
\begin{align}
I_m =  & \int_{\ell}\left(\prod_{i=1}^{m}G_{F}\left(q_{i}\right)\right)N\left(\ell,\{p_{k}\}\right)\,,
 \label{eq:numbcj1L}
\end{align}
where $\int_{\ell}=-\imath\int d^{d}\ell/\left(2\pi\right)^{d}\,,G_{F}\left(q_{i}\right)=\left(q_{i}^{2}+\imath0\right)^{-1}$
are massless Feynman propagators, and $N\left(\ell,\{p_{k}\}\right)$ is the
numerator built from the Jacobi off-shell currents $J_{\text{s,q,g}}$. 
The natural study of this object is considering its behaviour when it is embedded 
in one-loop topologies with two and three internal propagators, 
as depicted in Fig.~\ref{fig:Jac2pt1L}.
\begin{figure}[htb!]
\includegraphics[scale=0.85]{figs/fig3.epsi}
\caption{Jacobi combinations for $2\to 2$ processes at one-loop.}
\label{fig:Jac2pt1L}
\end{figure}

From Fig.~\ref{fig:Jac2pt1L} we note that at one-loop level we have an off-shell current attached to 
two off- and two on-shell momenta. Hence, we recall the results of Sec.~\ref{sec:embedding} for $N^{\mu_1\hdots\mu_4}_{\text{s,q,g}}$, whose decomposition in terms of off-shell momenta 
allows to provide relations at integral level. 
It turns out that if the numerator~\eqref{eq:numbcj1L} sits in three or two loop propagators, 
it collapses into 
\begin{subequations}
\begin{align}
I_3 &= \int_{\ell} \frac{N_3}{\ell^2(\ell+p_3)^2(\ell+p_{34})^2}\nn
&=\int_{\ell}\left\{
\tilde{A}_{11}\,\mathcal{I}\left[\!\!\!\parbox{25mm}{\input{figs/bubble_p1.tex}}\,\right] 
+ \tilde{A}_{12}\,\mathcal{I}\left[\!\!\!\parbox{25mm}{\input{figs/bubble_p2.tex}}\,\right]  
+ \tilde{C}_{11}\,\mathcal{I}\left[\parbox{15mm}{\input{figs/tadpole_p2341.tex}}\,\right]\right\} \,,
\end{align}
and
\begin{align}
I_2 &= \int_{\ell} \frac{N_2}{\ell^2(\ell+p_{34})^2}\nn
&= \int_{\ell}\left\{
 \tilde{A}_{21}\,\mathcal{I}\left[\parbox{15mm}{\input{figs/tadpole_p1234.tex}}\,\right] 
+ \tilde{A}_{22}\,\mathcal{I}\left[\parbox{15mm}{\input{figs/tadpole_p3412.tex}}\,\right] \,
+ \tilde{C}_{11}\,\mathcal{I}\left[\parbox{20mm}{\input{figs/contact_p1234.tex}}\,\right]\right\} \,.
\end{align}
\label{eq:I1I2}
\end{subequations}
Here, $\mathcal{I}[\cdots]$ states for the integrand of the topology, $\tilde{A}$ and $\tilde{C}$ are polynomials
in the loop momentum $\ell$, whose structure follows the results of Sec.~\ref{sec:embedding}. 

We observe that the r.h.s. of Eq.~\eqref{eq:I1I2} corresponds to topologies that integrate to zero in
dimensional regularisation, massless bubbles and tadpoles. 
Therefore, because of the Jacobi numerators~(\ref{eq:JsOffDec},\ref{eq:JqOffDec}) and~\eqref{eq:JgOffDec}, 
we obtain relations between 
integrals with the same number of loop propagators, three and two. 
These relations play a fundamental role when calculating an amplitude through the 
Loop-Tree duality (LTD) formalism. This is because different combinations of integrals 
can vanish by means of the outcome of CKD. We remark that apart of performing the single
cuts given by the prescriptions of LTD we also integrate out the expressions. 
It is indeed for this reason that the use of relations~\eqref{eq:I1I2} before 
integrating increases the efficiency in the evaluation of any amplitude. 
Since terms that seem to have a cumbersome structure at integrand level can be replaced  
by others holding a simpler structure. 

The recent work of Ref.s~\cite{Tourkine:2016bak,Ochirov:2017jby} has addressed an alternative study of numerators built from 
the off-shell current $J_{\text{g}}$. The approach of the authors is based on monodromy relations
in string theory. In particular, with the limit $\alpha'\to0$, the expansion of monodromy relations 
in terms of cubic diagrams turns out to be a linear combination of Jacobi identities. 

For the sake of simplicity, let us recall the results of Ref.~\cite[Eq.~(4.12)]{Ochirov:2017jby} for $2\to2$ processes,
which, in the notation of the current paper, become
\begin{align}
 0 =&\int_{\ell}\Bigg\{
\frac{1}{\ell^2 (\ell\!+\!p_{12})^2 (\ell\!-\!p_4)^2}n\!\left(\parbox{25mm}{\input{figs/bcj3_p1234.tex}}\,\right)
 - \frac{1}{\ell^2 (\ell\!+\!p_2)^2 (\ell\!+\!p_{23})^2}n\!\left(\parbox{25mm}{\input{figs/bcj3_p1423.tex}}\,\right)\nn
 &+ \frac{1}{\ell^2 (\ell\!+\!p_2)^2 (\ell\!-\!p_{4})^2}n\!\left(\parbox{25mm}{\input{figs/bcj3_p1324.tex}}\,\right)
  + \frac{1}{s_{12} \ell^2 (\ell\!+\!p_{12})^2}n\!\left(\parbox{30mm}{\input{figs/bcj2_p1234.tex}}\,\right)\nn
& - \frac{1}{s_{23} (\ell\!+\!p_1)^2 (\ell\!-\!p_4)^2}n\!\left(\parbox{30mm}{\input{figs/bcj2_p1423.tex}}\,\right)
 + \frac{1}{s_{24}(\ell\!+\!p_2)^2 (\ell\!-\!p_4)^2}n\!\left(\parbox{30mm}{\input{figs/bcj2_p1342.tex}}\,\right)
 \Bigg\}
 \,.
 \label{eq:4gMR}
\end{align}
The $+\imath 0$ prescription of the Feynman propagators is implicitly understood. 

From the diagrammatic and deep study of the Jacobi off-shell 
currents provided in Sec.s~\ref{sec:offcurrents} and~\ref{sec:decomposition}, 
we observe that the very same behaviour of Eq.~\eqref{eq:4gMR}
is recovered in decompositions~(\ref{eq:JsOffDec},\ref{eq:JqOffDec}) and~\eqref{eq:JgOffDec}. 
Moreover, the difference relies on the fact that we know 
\textit{apriori} the structure of any object built from the off-shell currents $J_{\text{s,q,g}}$. 
Therefore, we can eliminate topologies that at this point do not contribute and generate redundant 
contributions. This set of $2\to2$ massless processes do not show any difficulty in the calculation, 
since grouping integrands according to Jacobi identities satisfies CKD at integral level.

\subsection{$gg\to ss$ at one-loop}
\label{sec:exampleLTD}

In this section, we provide an application of CKD for one-loop $2\to 2$ processes. 
To this end, we embed the Jacobi off-shell currents in a one-loop topology. 
Since for this set up two particles are on- and the other two are off-shell, we consider,
without loose of generality, the off-shell current $J_{\text{g}}$. 
This is because the behaviour of the complete numerator $N_{\text{g}}$ will be the same as for 
$N_{\text{s}}$ and $N_{\text{q}}$ when two of the four particles attached to the Jacobi off-shell 
current are set on-shell, as shown in Sec.~\ref{sec:embedding}. 
This calculation is made in the framework of the LTD.

In order to write numerators as simple as posible 
we decompose them in terms of form factors by using Lorentz invariance.
In fact, the recent work of~\cite{Driencourt-Mangin:2017gop} deals with the 
Higgs production via gluon fusion, $gg\to H$ using LTD formalism. 
We can extend this set up for the current calculation.
This is due to the fact that the particles carrying 
the information of the process are  external gluons. 

Let us consider a one-loop object
\begin{align}
|\mathcal{M}_{gg\to ss}^{(1)}\rangle&=ig_{s}^{4}\,\varepsilon^{\mu_{1}}\left(p_{1}\right)\varepsilon^{\mu_{2}}\left(p_{2}\right)\mathcal{A}_{\mu_{1}\mu_{2}}^{(1)}\,,
\label{eq:ss2gg}
\end{align}
where $\mathcal{A}$ is built from the Jacobi off-shell current $J_{\text{g}}$. 
As mentioned above, because of Lorentz invariance, this tensor is given by
\begin{align}
\mathcal{A}_{\mu_{1}\mu_{2}}^{(1)} = \sum_{i=1}^{10} A_{i}^{(1)}T_{\mu_{1}\mu_{2}}^{i}\,,
\label{eq:decoAmp}
\end{align}
with $A_{i}$ the form factors or scalar functions of the Maldestam variables $s_{12},s_{23}$ and of the dimension $d$, 
and $T$ the tensor basis
\begin{align}
T_{i}^{\mu_{1}\mu_{2}}=&\Bigg\{ g^{\mu_{1}\mu_{2}}-\frac{2\,p_{1}^{\mu_{2}}p_{2}^{\mu_{1}}}{s_{12}}\,,g^{\mu_{1}\mu_{2}}\,,\frac{2\,p_{1}^{\mu_{2}}p_{3}^{\mu_{1}}}{s_{13}}\,,\frac{2\,p_{2}^{\mu_{1}}p_{3}^{\mu_{2}}}{s_{23}}\,,\frac{2\,s_{12}\,p_{3}^{\mu_{1}}p_{3}^{\mu_{2}}}{s_{23}s_{13}}\,,\nn
&\frac{2\,p_{1}^{\mu_{1}}p_{2}^{\mu_{2}}}{s_{12}}\,,\frac{2\,p_{1}^{\mu_{1}}p_{1}^{\mu_{2}}}{s_{12}}\,,\frac{2\,p_{2}^{\mu_{1}}p_{2}^{\mu_{2}}}{s_{12}}\,,\frac{2\,p_{1}^{\mu_{1}}p_{3}^{\mu_{2}}}{s_{23}}\,,\frac{2\,p_{2}^{\mu_{2}}p_{3}^{\mu_{1}}}{s_{13}}\Bigg\}\,,
\end{align}
with $s_{13}=-s_{12}-s_{23}$.

Since external gluons obey the transversality condition, $p_i\cdot\varepsilon(p_i)=0$, we only compute 
the first five contributions of~\eqref{eq:decoAmp}. These scalar coefficients are extracted 
by applying appropriate projecting operators on the tensor~\eqref{eq:decoAmp},
such that $P_i^{\mu\nu} {\cal A}^{(1)}_{\mu\nu} =  A_i^{(1)}$. 
The full set of projectors are summarised in Appendix~\ref{app:Projs}.
\\

In Sec.~\ref{sec:embedding} we studied the general case of the four off-shell legs attached to the 
Jacobi off-shell currents. For the one-loop case we are considering, with two off-shell particles, 
the parametric decomposition~\eqref{eq:JgOffDec} takes the form,
\begin{align}
N_{g}&=c_{2}\,\left(\ell+p_{34}\right)^{2}+c_{3}\,\ell^{2}+c_{23}\,\ell^{2}\left(\ell+p_{34}\right)^{2}\,,
\label{eq:Nex}
\end{align}
with $c_i$ scalar coefficients depending on the external and internal momenta, and $\ell$ the internal
loop momentum. Since this numerator is sat either in two or three loop massless propagators,
we can independently elaborate on each coefficient. 
This is done because each coefficient together with the set of propagators vanishes in 
dimensional regularisation as depicted in Eq.~\eqref{eq:I1I2}.\\

In order to explain how integral relations are obtained, let us first consider the 
case when the numerator~\eqref{eq:Nex} is 
sat in three loop propagators (see Fig.~\ref{fig:Jac2pt1L}.a). 
We find, for the form factor $A_1$ two independent identities\footnote{The scalar coefficients $c_{i}^{(j)}$ follow
the structure of the ones of Eq.~\eqref{eq:Nex} and account for the information of the projectors $P_{j}^{\mu\nu}$
of Eq.~\eqref{eq:projs}.}, 
\begin{subequations}
\begin{multline}
\int_{\ell}\frac{c_{2}^{(1)}\,\left(\ell+p_{34}\right)^{2}}{\ell^{2}(\ell+p_{3})^{2}(\ell+p_{34})^{2}}=\int_{\ell}\Bigg\{\delta(k_{1})\frac{\left(2k_{1}\cdot p_{1}+s_{23}\right)}{\left(2k_{1}\cdot p_{4}\right)}\Bigg[\left(k_{1}\cdot p_{1}\right)\Big(-\left(s_{12}^{2}+s_{12}s_{23}-s_{23}^{2}\right)\left(k_{1}\cdot p_{3}\right)\\+s_{23}\left(s_{12}^{2}+4s_{12}s_{23}+2s_{23}^{2}-s_{13}\left(k_{1}\cdot p_{4}\right)\right)+s_{12}s_{13}\left(k_{1}\cdot p_{2}\right)\Big)+s_{13}^{2}\left(k_{1}\cdot p_{2}\right)\left(k_{1}\cdot p_{34}+2s_{23}\right)-s_{12}s_{23}\left(k_{1}\cdot p_{1}\right){}^{2}\\+s_{12}\left(-s_{13}\left(k_{1}\cdot p_{3}\right){}^{2}+\left(k_{1}\cdot p_{3}\right)\left(-s_{13}\left(k_{1}\cdot p_{4}\right)+s_{23}(3s_{12}+4s_{23})\right)-s_{23}s_{13}\left(k_{1}\cdot p_{4}+s_{23}\right)\right)\Bigg]\\-\delta(k_{3})\frac{\left(k_{3}\cdot p_{1}+s_{23}\right)}{\left(2k_{3}\cdot p_{4}\right)}\Bigg[\left(k_{3}\cdot p_{1}\right)\Big(-2\left(s_{12}^{2}+s_{12}s_{23}-s_{23}^{2}\right)\left(k_{3}\cdot p_{3}\right)\\+2s_{12}s_{13}\left(k_{3}\cdot p_{2}\right)+2s_{23}s_{13}\left(k_{3}\cdot p_{4}\right)+s_{23}(s_{12}+2s_{23})(3s_{12}+2s_{23})\Big)\\-2s_{12}s_{23}\left(k_{3}\cdot p_{1}\right){}^{2}+\left(-s_{13}\left(k_{3}\cdot p_{2}\right)+s_{12}\left(k_{3}\cdot p_{3}\right)\right)\left(-2s_{13}k_{3}\cdot\left(p_{3}-p_{4}\right)+(s_{12}+2s_{23})^{2}\right)\Bigg]\Bigg\}=0\,,
\label{eq:A1eqA}
\end{multline}
\begin{multline}
\int_{\ell}\frac{c_{3}^{(1)}\,\ell^{2}}{\ell^{2}(\ell+p_{3})^{2}(\ell+p_{34})^{2}}=
\int_{\ell}\Bigg\{\frac{\tilde{\delta}(k_{1})\left(2k_{1}\cdot p_{1}-s_{13}\right)}{8\left(2k_{1}\cdot p_{3}\right)}\Bigg[\left(k_{1}\cdot p_{1}\right)\Big(2s_{12}s_{13}\left(k_{1}\cdot p_{2}\right)+2s_{23}s_{13}\left(k_{1}\cdot p_{4}\right)\\-2\left(s_{12}^{2}+3s_{23}s_{12}+s_{23}^{2}\right)\left(k_{1}\cdot p_{3}\right)-s_{23}\left(s_{12}+2s_{23}\right)\left(3s_{12}+2s_{23}\right)\Big)\\-2s_{12}s_{23}\left(k_{1}\cdot p_{1}\right){}^{2}+s_{13}\left(k_{1}\cdot p_{2}\right)\left(\left(s_{12}+2s_{23}\right){}^{2}-2s_{13}\left(k_{1}\cdot p_{3}+k_{1}\cdot p_{4}\right)\right)\\+s_{12}\left(-\left(k_{1}\cdot p_{3}\right)\left(-2s_{13}k_{1}\cdot p_{34}+s_{12}^{2}+2s_{23}s_{12}+2s_{23}^{2}\right)-s_{13}s_{23}\left(s_{12}+2s_{23}\right)\right)\Bigg]\\+\frac{\tilde{\delta}(k_{2})\left(k_{2}\cdot p_{1}-s_{13}\right)}{4\left(2k_{2}\cdot p_{3}\right)}\Bigg[\left(k_{2}\cdot p_{1}\right)\Big(-2s_{12}s_{13}\left(k_{2}\cdot p_{2}\right)-2s_{23}s_{13}\left(k_{2}\cdot p_{4}\right)\\+2\left(s_{12}^{2}+s_{23}s_{12}-s_{23}^{2}\right)\left(k_{2}\cdot p_{3}\right)+s_{23}\left(s_{12}+2s_{23}\right)\left(3s_{12}+2s_{23}\right)\Big)\\+2s_{12}s_{23}\left(k_{2}\cdot p_{1}\right){}^{2}+\left(s_{12}\left(k_{2}\cdot p_{3}\right)-s_{13}\left(k_{2}\cdot p_{2}\right)\right)\left(2s_{13}k_{2}\cdot\left(p_{3}-k_{2}\right)+\left(s_{12}+2s_{23}\right){}^{2}\right)\Bigg]\Bigg\}=0\,,
\label{eq:A1eqB}
\end{multline}
\label{eq:IntRel1}
\end{subequations}
Terms proportional to $\ell^{2}\left(\ell+p_{34}\right)^{2}$ do not appear in this analysis. 

Similarly for $A_2$, 
\begin{subequations}
\begin{multline}
\int_{\ell}\frac{c_{2}^{(2)}\,\left(\ell+p_{34}\right)^{2}}{\ell^{2}(\ell+p_{3})^{2}(\ell+p_{34})^{2}}=
\int_{\ell}\Bigg\{\frac{\tilde{\delta}(k_{1})\left(2\left(k_{1}\cdot p_{1}\right)+s_{23}\right)}{\left(2k_{1}\cdot p_{4}\right)}\Bigg[\left(k_{1}\cdot p_{1}\right)\Big(-s_{13}s_{23}\left(k_{1}\cdot p_{4}\right)+s_{12}s_{13}\left(k_{1}\cdot p_{2}\right)\\-\left(s_{12}^{2}+s_{23}s_{12}-s_{23}^{2}\right)\left(k_{1}\cdot p_{3}\right)+s_{12}s_{23}^{2}\Big)\\-s_{12}s_{23}\left(k_{1}\cdot p_{1}\right){}^{2}-s_{13}\left(k_{1}\cdot p_{2}\right)\left(s_{12}s_{23}-s_{13}\left(k_{1}\cdot p_{34}\right)\right)\\+s_{12}\left(-s_{13}\left(k_{1}\cdot p_{4}\right)\left(k_{1}\cdot p_{3}+2s_{23}\right)+\left(k_{1}\cdot p_{3}\right)\left(s_{23}\left(4s_{12}+5s_{23}\right)-s_{13}\left(k_{1}\cdot p_{3}\right)\right)+s_{12}s_{13}s_{23}\right)\Bigg]\\+\frac{\tilde{\delta}(k_{3})\left(k_{3}\cdot p_{1}+s_{23}\right)}{\left(2k_{3}\cdot p_{4}\right)}\Bigg[\left(k_{3}\cdot p_{1}\right)\Big(2s_{12}s_{13}\left(k_{3}\cdot p_{2}\right)+2s_{23}s_{13}\left(k_{3}\cdot p_{4}\right)\\-2\left(s_{12}^{2}+s_{23}s_{12}-s_{23}^{2}\right)\left(k_{3}\cdot p_{3}\right)+s_{12}s_{23}\left(s_{12}+2s_{23}\right)\Big)\\+2s_{12}s_{13}\left(k_{3}\cdot p_{4}\right)\left(k_{3}\cdot p_{3}+s_{23}\right)+s_{12}\left(k_{3}\cdot p_{3}\right)\left(-2s_{13}\left(k_{3}\cdot p_{3}\right)+s_{12}^{2}+6s_{23}s_{12}+6s_{23}^{2}\right)\\-2s_{12}s_{23}\left(k_{3}\cdot p_{1}\right){}^{2}-s_{13}\left(k_{3}\cdot p_{2}\right)\left(s_{12}\left(s_{12}+2s_{23}\right)-2s_{13}k_{3}\cdot\left(p_{3}-p_{4}\right)\right)\Bigg]\Bigg\}=0\,,
\end{multline}
\begin{multline}
\int_{\ell}\frac{c_{3}^{(2)}\,\ell^{2}}{\ell^{2}(\ell+p_{3})^{2}(\ell+p_{34})^{2}}=
\int_{\ell}\Bigg\{\frac{\tilde{\delta}(k_{1})\left(2\left(k_{1}\cdot p_{1}\right)-s_{13}\right)}{2\left(2k_{1}\cdot p_{3}\right)}\Bigg[\left(k_{1}\cdot p_{1}\right)\Big(2s_{12}s_{13}\left(k_{1}\cdot p_{2}\right)+2s_{23}s_{13}\left(k_{1}\cdot p_{4}\right)\\-2\left(s_{12}^{2}+3s_{23}s_{12}+s_{23}^{2}\right)\left(k_{1}\cdot p_{3}\right)-s_{12}s_{23}\left(s_{12}+2s_{23}\right)\Big)\\+s_{12}\left(2s_{13}\left(k_{1}\cdot p_{3}\right){}^{2}-\left(k_{1}\cdot p_{3}\right)\left(s_{12}^{2}-2s_{13}\left(k_{1}\cdot p_{4}\right)\right)-s_{13}s_{23}\left(2\left(k_{1}\cdot p_{4}\right)+s_{12}\right)\right)\\-2s_{12}s_{23}\left(k_{1}\cdot p_{1}\right){}^{2}+s_{13}\left(k_{1}\cdot p_{2}\right)\left(s_{12}\left(s_{12}+2s_{23}\right)-2s_{13}\left(k_{1}\cdot p_{3}+k_{1}\cdot p_{4}\right)\right)\Bigg]\\+\frac{\delta(k_{2})\left(k_{2}\cdot p_{1}-s_{13}\right)}{\left(2k_{2}\cdot p_{3}\right)}\Bigg[\left(k_{2}\cdot p_{1}\right)\Big(-2s_{12}s_{13}\left(k_{2}\cdot p_{2}\right)-2s_{23}s_{13}\left(k_{2}\cdot p_{4}\right)\\+2\left(s_{12}^{2}+s_{23}s_{12}-s_{23}^{2}\right)\left(k_{2}\cdot p_{3}\right)+s_{12}s_{23}\left(s_{12}+2s_{23}\right)\Big)\\-2s_{12}s_{13}\left(k_{2}\cdot p_{4}\right)\left(k_{2}\cdot p_{3}-s_{23}\right)+s_{12}\left(k_{2}\cdot p_{3}\right)\left(2s_{13}\left(k_{2}\cdot p_{3}\right)+s_{12}^{2}+6s_{23}s_{12}+6s_{23}^{2}\right)\\+2s_{12}s_{23}\left(k_{2}\cdot p_{1}\right){}^{2}-s_{13}\left(k_{2}\cdot p_{2}\right)\left(2s_{13}\left(k_{2}\cdot p_{3}-k_{2}\cdot p_{4}\right)+s_{12}\left(s_{12}+2s_{23}\right)\right)\Bigg]\Bigg\}=0\,,
\end{multline}
\label{eq:IntRel2}
\end{subequations}
where $\tilde{\delta}(k_i) = 2\pi\,\imath\,\delta^{(+)}(k_i^2)$ and  $k_i$ are defined as follows, 
\begin{align}
k_{1}=\ell+p_{4}\,,&&k_{2}=\ell+p_{34}\,,&&k_{3}=\ell\,,
\end{align}
whose on-shell loop energies are given by 
\begin{align}
k_{1,0}^{\left(+\right)}=\sqrt{\left(\boldsymbol{\ell}+\boldsymbol{p_{4}}\right)^{2}},&&k_{2,0}^{\left(+\right)}=\sqrt{\left(\boldsymbol{\ell}+\boldsymbol{p_{34}}\right)^{2}},&&k_{3,0}^{\left(+\right)}=\sqrt{\boldsymbol{\ell}^{2}}\,,.
\end{align}
In Eq.s~\eqref{eq:IntRel1} and~\eqref{eq:IntRel2} we observe relations between integrals that appear from different
cuts. Nevertheless, for these $2\to 2$ processes it is convenient to work in the center-of-mass frame, 
because of an additional constraint $k_{2,0}^{\left(+\right)}=k_{3,0}^{\left(+\right)}=\ell_{0}^{\left(+\right)}$, 
$\tilde{\delta}(k_2)=\tilde{\delta}(k_3)=\tilde{\delta}(\ell)$.
This constraint  allows us to combine both results, for instance the two identities in 
Eq.s~\eqref{eq:IntRel1} or~\eqref{eq:IntRel2}, to express the integrals that depend on $\tilde{\delta}(k_1)$ in terms
of simpler ones. 

A similar study, even though it is not that illuminating for the purpose of this example, 
can be done for the other form factors, $A_i,i=3,4,5$, as well as for 
the case of the numerator~\eqref{eq:Nex}
with two loop propagators, which is depicted in Fig.~\ref{fig:Jac2pt1L}.b. 
The difference with the above example relies on the rank of the numerators. 
Since for this one we obtain a polynomial in $\ell$ with at most rank two.
However, in order to get further simplifications in the integral relations,
contributions of~\ref{fig:Jac2pt1L}.b do have to be taken into account.

%% file: figs/bubble_p1.tex
\begin{tikzpicture}[line width=1 pt,node distance=0.5 cm and 0.5 cm]
\coordinate[] (v1);
\coordinate[left = of v1, label= left :\footnotesize$p_1$] (p1);
\coordinate[right = of v1] (v2);
\coordinate[above right = of v2, label= right :\footnotesize$p_2$] (p2);
\coordinate[right = of v2, label= right :\footnotesize$p_3$] (p3);
\coordinate[below right = of v2, label= right :\footnotesize$p_4$] (p4);

\draw[fermionnoarrow] (p1) -- (v1);
\semiloop[fermion]{v1}{v2}{0};
\semiloop[fermionnoarrow]{v2}{v1}{180};
\draw[fermionnoarrow] (p2) -- (v2);
\draw[fermionnoarrow] (p3) -- (v2);
\draw[fermionnoarrow] (p4) -- (v2);

\end{tikzpicture}

%% file: figs/bubble_p2.tex
\begin{tikzpicture}[line width=1 pt,node distance=0.5 cm and 0.5 cm]
\coordinate[] (v1);
\coordinate[left = of v1, label= left :\footnotesize$p_2$] (p1);
\coordinate[right = of v1] (v2);
\coordinate[above right = of v2, label= right :\footnotesize$p_3$] (p2);
\coordinate[right = of v2, label= right :\footnotesize$p_4$] (p3);
\coordinate[below right = of v2, label= right :\footnotesize$p_1$] (p4);

\draw[fermionnoarrow] (p1) -- (v1);
\semiloop[fermionnoarrow]{v1}{v2}{0};
\semiloop[fermion]{v2}{v1}{180};
\draw[fermionnoarrow] (p2) -- (v2);
\draw[fermionnoarrow] (p3) -- (v2);
\draw[fermionnoarrow] (p4) -- (v2);

\end{tikzpicture}

%% file: figs/tadpole_p2341.tex
\begin{tikzpicture}[line width=1 pt,node distance=0.25 cm and 0.5 cm]
\coordinate[] (v1);
\coordinate[right = of v1] (v2);
\coordinate[right = of v2] (v3);

\coordinate[above = of v3, label= right :\footnotesize$p_3$] (p3);
\coordinate[above = of p3, label= right :\footnotesize$p_2$] (p2);
\coordinate[below = of v3, label= right :\footnotesize$p_4$] (p4);
\coordinate[below = of p4, label= right :\footnotesize$p_1$] (p1);

\semiloop[fermionnoarrow]{v1}{v2}{0};
\semiloop[fermion]{v2}{v1}{180};
\draw[fermionnoarrow] (p1) -- (v2);
\draw[fermionnoarrow] (p2) -- (v2);
\draw[fermionnoarrow] (p3) -- (v2);
\draw[fermionnoarrow] (p4) -- (v2);

\end{tikzpicture}

%% file: figs/tadpole_p1234.tex
\begin{tikzpicture}[line width=1 pt,node distance=0.25 cm and 0.5 cm]
\coordinate[] (v1);
\coordinate[right = of v1] (v2);
\coordinate[right = of v2] (v3);

\coordinate[above = of v3, label= right :\footnotesize$p_2$] (p3);
\coordinate[below = of v3, label= right :\footnotesize$p_3$] (p4);
\coordinate[below = of p4, label= right :\footnotesize$p_4$] (p1);
\coordinate[above = of p3, label= right :\footnotesize$p_1$] (p2);

\semiloop[fermionnoarrow]{v1}{v2}{0};
\semiloop[fermion]{v2}{v1}{180};
\draw[fermionnoarrow] (p1) -- (v2);
\draw[fermionnoarrow] (p2) -- (v2);
\draw[fermionnoarrow] (p3) -- (v2);
\draw[fermionnoarrow] (p4) -- (v2);

\end{tikzpicture}

%% file: figs/tadpole_p3412.tex
\begin{tikzpicture}[line width=1 pt,node distance=0.25 cm and 0.5 cm]
\coordinate[] (v1);
\coordinate[right = of v1] (v2);
\coordinate[right = of v2] (v3);

\coordinate[above = of v3, label= right :\footnotesize$p_4$] (p3);
\coordinate[below = of v3, label= right :\footnotesize$p_1$] (p4);
\coordinate[below = of p4, label= right :\footnotesize$p_2$] (p1);
\coordinate[above = of p3, label= right :\footnotesize$p_3$] (p2);

\semiloop[fermionnoarrow]{v1}{v2}{0};
\semiloop[fermion]{v2}{v1}{180};
\draw[fermionnoarrow] (p1) -- (v2);
\draw[fermionnoarrow] (p2) -- (v2);
\draw[fermionnoarrow] (p3) -- (v2);
\draw[fermionnoarrow] (p4) -- (v2);

\end{tikzpicture}

%% file: figs/contact_p1234.tex
\begin{tikzpicture}[line width=1 pt,node distance=0.25 cm and 0.5 cm]
\coordinate[] (v1);

\coordinate[above left = of v1, label= left :\footnotesize$p_4$] (p4);
\coordinate[below left = of v1, label= left :\footnotesize$p_3$] (p3);
\coordinate[below right = of v1, label= right :\footnotesize$p_2$] (p2);
\coordinate[above right = of v1, label= right :\footnotesize$p_1$] (p1);

\draw[fermionnoarrow] (p1) -- (v1);
\draw[fermionnoarrow] (p2) -- (v1);
\draw[fermionnoarrow] (p3) -- (v1);
\draw[fermionnoarrow] (p4) -- (v1);

\end{tikzpicture}

%% file: figs/bcj3_p1234.tex
\begin{tikzpicture}[line width=1 pt,node distance=0.5 cm and 0.5 cm]

\coordinate[] (v1);
\coordinate[below left = of v1, label= left :\footnotesize$p_1$] (y1);
\coordinate[above left = of v1, label= left :\footnotesize$p_2$] (x1);
\coordinate[above right = of v1] (x2);
\coordinate[below right = of v1] (y2);

\draw[fermionnoarrow] (x2) -- (y2);
\draw[fermionnoarrow] (v1) -- (x2);
\draw[fermionnoarrow] (v1) -- (y2);

\coordinate[ right = of x2, label= right :\footnotesize$p_3$] (p2);
\coordinate[ right = of y2, label= right :\footnotesize$p_4$] (p3);

\draw[fermionnoarrow] (x2) -- (p2);
\draw[fermionnoarrow] (y2) -- (p3);
\draw[fermionnoarrow] (v1) -- (x1);
\draw[fermionnoarrow] (v1) -- (y1);

\draw[fill=black] (x2) circle (.05cm);
\draw[fill=black] (y2) circle (.05cm);
\draw[fill=white] [pos=0.5] circle (.35cm);
\draw node [pos=0.5] {$\boldsymbol{J}$};
\end{tikzpicture}

%% file: figs/bcj3_p1423.tex
\begin{tikzpicture}[line width=1 pt,node distance=0.5 cm and 0.5 cm]

\coordinate[] (v1);
\coordinate[above left = of v1, label= left :\footnotesize$p_4$] (x1);
\coordinate[below left = of v1, label= left :\footnotesize$p_1$] (y1);
\coordinate[above right = of v1] (x2);
\coordinate[below right = of v1] (y2);

\draw[fermionnoarrow] (x2) -- (y2);
\draw[fermionnoarrow] (v1) -- (x2);
\draw[fermionnoarrow] (v1) -- (y2);

\coordinate[ right = of x2, label= right :\footnotesize$p_2$] (p2);
\coordinate[ right = of y2, label= right :\footnotesize$p_3$] (p3);

\draw[fermionnoarrow] (x2) -- (p2);
\draw[fermionnoarrow] (y2) -- (p3);
\draw[fermionnoarrow] (v1) -- (x1);
\draw[fermionnoarrow] (v1) -- (y1);

\draw[fill=black] (x2) circle (.05cm);
\draw[fill=black] (y2) circle (.05cm);
\draw[fill=white] [pos=0.5] circle (.35cm);
\draw node [pos=0.5] {$\boldsymbol{J}$};
\end{tikzpicture}

%% file: figs/bcj3_p1324.tex
\begin{tikzpicture}[line width=1 pt,node distance=0.5 cm and 0.5 cm]

\coordinate[] (v1);
\coordinate[above left = of v1, label= left :\footnotesize$p_3$] (x1);
\coordinate[below left = of v1, label= left :\footnotesize$p_1$] (y1);
\coordinate[above right = of v1] (x2);
\coordinate[below right = of v1] (y2);

\draw[fermionnoarrow] (x2) -- (y2);
\draw[fermionnoarrow] (v1) -- (x2);
\draw[fermionnoarrow] (v1) -- (y2);

\coordinate[ right = of x2, label= right :\footnotesize$p_4$] (p2);
\coordinate[ right = of y2, label= right :\footnotesize$p_2$] (p3);

\draw[fermionnoarrow] (x2) -- (p2);
\draw[fermionnoarrow] (y2) -- (p3);
\draw[fermionnoarrow] (v1) -- (x1);
\draw[fermionnoarrow] (v1) -- (y1);

\draw[fill=black] (x2) circle (.05cm);
\draw[fill=black] (y2) circle (.05cm);
\draw[fill=white] [pos=0.5] circle (.35cm);
\draw node [pos=0.5] {$\boldsymbol{J}$};
\end{tikzpicture}

%% file: figs/bcj2_p1234.tex
\begin{tikzpicture}[line width=1 pt,node distance=0.5 cm and 0.5 cm]

\coordinate[] (v1);
\coordinate[above left = of v1, label= left :\footnotesize$p_2$] (x1);
\coordinate[below left = of v1, label= left :\footnotesize$p_1$] (y1);
\coordinate[ right = of v1] (v2);
\coordinate[ right = of v2] (x3);
\coordinate[above right = of x3, label= right :\footnotesize$p_3$] (p2);
\coordinate[below right = of x3, label= right :\footnotesize$p_4$] (p3);
\coordinate[below left = of v1] (p4);
\coordinate[above left = of v1] (p1);

\semiloop[fermionnoarrow]{v1}{v2}{0};
\semiloop[fermionnoarrow]{v2}{v1}{180};

\draw[fermionnoarrow] (v2) -- (x3);
\draw[fermionnoarrow] (x3) -- (p2);
\draw[fermionnoarrow] (x3) -- (p3);
\draw[fermionnoarrow] (v1) -- (p1);
\draw[fermionnoarrow] (v1) -- (p4);

%
%
%
%
\draw[fill=black] (v2) circle (.05cm);
\draw[fill=black] (x3) circle (.05cm);
\draw[fill=white] [pos=0.5] circle (.35cm);
\draw node [pos=0.5] {$\boldsymbol{J}$};
\end{tikzpicture}

%% file: figs/bcj2_p1423.tex
\begin{tikzpicture}[line width=1 pt,node distance=0.5 cm and 0.5 cm]

\coordinate[] (v1);
\coordinate[above left = of v1, label= left :\footnotesize$p_4$] (x1);
\coordinate[below left = of v1, label= left :\footnotesize$p_1$] (y1);
\coordinate[ right = of v1] (v2);
\coordinate[ right = of v2] (x3);
\coordinate[above right = of x3, label= right :\footnotesize$p_2$] (p2);
\coordinate[below right = of x3, label= right :\footnotesize$p_3$] (p3);
\coordinate[below left = of v1] (p4);
\coordinate[above left = of v1] (p1);

\semiloop[fermionnoarrow]{v1}{v2}{0};
\semiloop[fermionnoarrow]{v2}{v1}{180};

\draw[fermionnoarrow] (v2) -- (x3);
\draw[fermionnoarrow] (x3) -- (p2);
\draw[fermionnoarrow] (x3) -- (p3);
\draw[fermionnoarrow] (v1) -- (p1);
\draw[fermionnoarrow] (v1) -- (p4);

%
%
%
%
\draw[fill=black] (v2) circle (.05cm);
\draw[fill=black] (x3) circle (.05cm);
\draw[fill=white] [pos=0.5] circle (.35cm);
\draw node [pos=0.5] {$\boldsymbol{J}$};
\end{tikzpicture}

%% file: figs/bcj2_p1342.tex
\begin{tikzpicture}[line width=1 pt,node distance=0.5 cm and 0.5 cm]

\coordinate[] (v1);
\coordinate[above left = of v1, label= left :\footnotesize$p_3$] (x1);
\coordinate[below left = of v1, label= left :\footnotesize$p_1$] (y1);
\coordinate[ right = of v1] (v2);
\coordinate[ right = of v2] (x3);
\coordinate[above right = of x3, label= right :\footnotesize$p_4$] (p2);
\coordinate[below right = of x3, label= right :\footnotesize$p_2$] (p3);
\coordinate[below left = of v1] (p4);
\coordinate[above left = of v1] (p1);

\semiloop[fermionnoarrow]{v1}{v2}{0};
\semiloop[fermionnoarrow]{v2}{v1}{180};

\draw[fermionnoarrow] (v2) -- (x3);
\draw[fermionnoarrow] (x3) -- (p2);
\draw[fermionnoarrow] (x3) -- (p3);
\draw[fermionnoarrow] (v1) -- (p1);
\draw[fermionnoarrow] (v1) -- (p4);

%
%
%
%
\draw[fill=black] (v2) circle (.05cm);
\draw[fill=black] (x3) circle (.05cm);
\draw[fill=white] [pos=0.5] circle (.35cm);
\draw node [pos=0.5] {$\boldsymbol{J}$};
\end{tikzpicture}

%% file: conclusions.tex
\section{Conclusions}

In this paper, we have studied the colour-kinematics duality (CKD) at tree and one-loop level. 
We have computed the off-shell currents in the Feynman and axial gauges
from the Jacobi-like combinations of numerators. We have considered the QCD processes 
$gg\to ss,q\bar{q},gg$, finding that the most compact form for these off-shell currents can always
be written in terms of three-point Feynman rules. Although, these Feynman rules do not obey momentum 
conservation they allow for a universal representation in terms of three-point interactions. 
We have also seen that CKD is satisfied when the four external particles
are set on-shell. 

Consequently, we have embedded the off-shell currents in a richer topology, at tree level with higher-multiplicity
or one-loop, obtaining that with a decomposition of  the four momenta 
entering to the off-shell currents in terms of on-shell massless ones, any object built from these currents 
can be written as a linear combination of at most two of the four squared momenta, $p_i^2p_j^2$. 
This outcome has been achieved as a byproduct of the choice of the gauge, since the reference momentum
has chosen to be the same for all internal gluons. 

The decomposition of one-loop numerators has been extended, finding relations between integrals with the same number 
of propagators. For the $2\to2$ case, we have written relations for Feynman integrals 
with two and three loop propagators. In the same manner and in the framework of the Loop-Tree duality
formalism, we have considered the particular example
of $gg\to ss$ showing that higher rank numerators can be replaced by lower ones. 
This outcome indeed allows for an optimisation in the evaluation of  Feynman integrals.

Furthermore, we have observed that monodromy relations obtained in Ref.~\cite{Ochirov:2017jby}
in tandem with the structure of numerators built from the Jacobi off-shell currents play a very important role for finding 
integral relations for $2\to n$ processes with $n>2$. In particular, the linear combination of Jacobi identities 
in the kinematical sector, provided by monodromy relations, generates all posible integral relations. 
However, several topologies turn out to be redundant, since the Jacobi identity of kinematic numerators
shows the same pattern we have discussed through this paper in which their contribution vanish upon integration. 

Future work could include the consequences of these integral relations at higher loop orders, 
in particular if the calculation of Integration-by-parts identities~\cite{Tkachov:1981wb,Chetyrkin:1981qh,Laporta:2001dd} gets optimised when there is a new ingredient provided by CKD. 

%% file: ackno.tex
\section*{Acknowledgments}
This work is supported by the Spanish
Government and ERDF funds from European Commission (Grants No. FPA2014-53631-C2-1-P and SEV-2014-
0398) and by Consejo Superior de Investigaciones Cient\'{\i}ficas (Grant No. PIE-201750E021). The work of J.L. has been supported by Instituto de F\'{\i}sica Corpuscular.

%% file: appFR.tex
\section{Three-point interaction Feynman rules}
\label{app:FRs}
In this appendix, we collect the colour stripped Feynman rules of Sec.~\ref{sec:preliminaries},
\begin{align}
\parbox{20mm}{\input{figs/GSS1.tex}} & = \left(p_{2}-p_{3}\right)^{\mu_{1}}\,,\label{eq:gss}\\[-0.5cm]
\parbox{20mm}{\input{figs/FFG1.tex}} &= \gamma^{\mu_1}\,,\label{eq:gqq}\\[-0.5cm]
\parbox{20mm}{\input{figs/FGG2.tex}} &= \frac{1}{2}[\gamma^{\mu_4},\gamma^{\mu_1}]\,,\label{eq:qgg}\\[-0.5cm]
\parbox{20mm}{\input{figs/GGG123.tex}} & = g^{\mu_{1}\mu_{2}}\left(p_{1}-p_{2}\right)^{\mu_{3}}+g^{\mu_{2}\mu_{3}}\left(p_{2}-p_{3}\right)^{\mu_{1}}+g^{\mu_{3}\mu_{1}}\left(p_{3}-p_{1}\right)^{\mu_{2}}\,.\label{eq:ggg}
\end{align}

%% file: figs/GSS1.tex
\unitlength=0.20bp%
\begin{feynartspicture}(300,300)(1,1)
\FADiagram{}
\FAProp(3.,10.)(10.,10.)(0.,){/Cycles}{0}
\FAProp(16.,15.)(10.,10.)(0.,){/ScalarDash}{0}
\FAProp(16.,5.)(10.,10.)(0.,){/ScalarDash}{0}
\FAVert(10.,10.){0}
\FALabel(16.2273,15.5749)[cb]{\tiny $p_2$}
\FALabel(16.2273,4.5749)[ct]{\tiny $p_3$}
\FALabel(2.3,8.93)[ct]{\tiny $p_1, \mu_1$}
\end{feynartspicture}

%% file: figs/FFG1.tex
\unitlength=0.20bp%
\begin{feynartspicture}(300,300)(1,1)
\FADiagram{}
\FAProp(3.,10.)(10.,10.)(0.,){/Cycles}{0}
\FAProp(16.,15.)(10.,10.)(0.,){/Straight}{-1}
\FAProp(16.,5.)(10.,10.)(0.,){/Straight}{1}
\FAVert(10.,10.){0}
\FALabel(16.2273,15.5749)[cb]{\tiny $p_2$}
\FALabel(16.2273,4.5749)[ct]{\tiny $p_3$}
\FALabel(2.3,8.93)[ct]{\tiny $p_1, \mu_1$}
\end{feynartspicture}

%% file: figs/FGG2.tex
\unitlength=0.20bp%
\begin{feynartspicture}(300,300)(1,1)
\FADiagram{}
\FAProp(3.,10.)(10.,10.)(0.,){/Straight}{-1}
\FAProp(16.,15.)(10.,10.)(0.,){/Cycles}{0}
\FAProp(16.,5.)(10.,10.)(0.,){/Cycles}{0}
\FAVert(10.,10.){0}
\FALabel(16.2273,15.5749)[cb]{\tiny $p_4, \mu_4$}
\FALabel(16.2273,4.5749)[ct]{\tiny $p_1, \mu_1$}
\FALabel(2.3,8.93)[ct]{\tiny $p_2$}
\end{feynartspicture}

%% file: figs/GGG123.tex
\unitlength=0.20bp%
\begin{feynartspicture}(300,300)(1,1)
\FADiagram{}
\FAProp(3.,10.)(10.,10.)(0.,){/Cycles}{0}
\FAProp(16.,15.)(10.,10.)(0.,){/Cycles}{0}
\FAProp(16.,5.)(10.,10.)(0.,){/Cycles}{0}
\FALabel(16.2273,15.5749)[cb]{\tiny $p_1, \mu_1$}
\FALabel(16.2273,4.5749)[ct]{\tiny $p_2, \mu_2$}
\FALabel(2.3,8.93)[ct]{\tiny $p_3, \mu_3$}
\FAVert(10.,10.){0}
\end{feynartspicture}

%% file: appProj.tex
\section{Projectors}
\label{app:Projs}

In this appendix, we collect the projectors used to extract the form
factors $A_{i}$ of Sec.~\ref{sec:exampleLTD}
\begin{subequations}
\begin{align}
P_{1}^{\mu_{1}\mu_{2}} & =\frac{1}{2\left(d-3\right)}\Bigg[g^{\mu_{1}\mu_{2}}-\frac{(d-2)p_{1}^{\mu_{2}}p_{2}^{\mu_{1}}}{s_{12}}+\frac{(d-4)p_{3}^{\mu_{2}}p_{2}^{\mu_{1}}}{s_{23}}+\frac{(d-2)s_{13}p_{2}^{\mu_{2}}p_{2}^{\mu_{1}}}{s_{12}s_{23}}+\frac{(d-2)s_{23}p_{1}^{\mu_{1}}p_{1}^{\mu_{2}}}{s_{12}s_{13}}\nonumber \\
 & \qquad-\frac{(d-2)p_{1}^{\mu_{1}}p_{2}^{\mu_{2}}}{s_{12}}+\frac{(d-4)p_{1}^{\mu_{2}}p_{3}^{\mu_{1}}}{s_{13}}-\frac{(d-2)p_{1}^{\mu_{1}}p_{3}^{\mu_{2}}}{s_{13}}-\frac{(d-2)p_{2}^{\mu_{2}}p_{3}^{\mu_{1}}}{s_{23}}-\frac{(d-4)s_{12}p_{3}^{\mu_{1}}p_{3}^{\mu_{2}}}{s_{13}s_{23}}\Bigg]\,,\\
P_{2}^{\mu_{1}\mu_{2}} & =\frac{1}{2\left(d-3\right)}\Bigg[g^{\mu_{1}\mu_{2}}+\frac{(d-4)p_{1}^{\mu_{2}}p_{2}^{\mu_{1}}}{s_{12}}-\frac{(d-2)p_{3}^{\mu_{2}}p_{2}^{\mu_{1}}}{s_{23}}-\frac{(d-4)s_{13}p_{2}^{\mu_{2}}p_{2}^{\mu_{1}}}{s_{12}s_{23}}-\frac{(d-4)s_{23}p_{1}^{\mu_{1}}p_{1}^{\mu_{2}}}{s_{12}s_{13}}\nonumber \\
 & \qquad+\frac{(d-4)p_{1}^{\mu_{1}}p_{2}^{\mu_{2}}}{s_{12}}-\frac{(d-2)p_{1}^{\mu_{2}}p_{3}^{\mu_{1}}}{s_{13}}+\frac{(d-4)p_{1}^{\mu_{1}}p_{3}^{\mu_{2}}}{s_{13}}+\frac{(d-4)p_{2}^{\mu_{2}}p_{3}^{\mu_{1}}}{s_{23}}+\frac{(d-2)s_{12}p_{3}^{\mu_{1}}p_{3}^{\mu_{2}}}{s_{13}s_{23}}\Bigg]\,,\\
P_{3}^{\mu_{1}\mu_{2}} & =\frac{1}{\left(d-3\right)}\Bigg[g^{\mu_{1}\mu_{2}}-\frac{(d-2)p_{1}^{\mu_{2}}p_{2}^{\mu_{1}}}{s_{12}}-\frac{(d-2)p_{3}^{\mu_{2}}p_{2}^{\mu_{1}}}{s_{23}}+\frac{(d-2)s_{13}p_{2}^{\mu_{2}}p_{2}^{\mu_{1}}}{s_{12}s_{23}}+\frac{(d-2)s_{23}p_{1}^{\mu_{1}}p_{1}^{\mu_{2}}}{s_{12}s_{13}}\nonumber \\
 & \qquad-\frac{(d-2)p_{1}^{\mu_{1}}p_{2}^{\mu_{2}}}{s_{12}}-\frac{(d-2)p_{1}^{\mu_{2}}p_{3}^{\mu_{1}}}{s_{13}}+\frac{(d-4)p_{1}^{\mu_{1}}p_{3}^{\mu_{2}}}{s_{13}}+\frac{(d-4)p_{2}^{\mu_{2}}p_{3}^{\mu_{1}}}{s_{23}}-\frac{(d-4)s_{12}p_{3}^{\mu_{1}}p_{3}^{\mu_{2}}}{s_{13}s_{23}}\Bigg]\,,\\
P_{4}^{\mu_{1}\mu_{2}} & =-\frac{1}{\left(d-3\right)}\Bigg[g^{\mu_{1}\mu_{2}}-\frac{(d-2)p_{1}^{\mu_{2}}p_{2}^{\mu_{1}}}{s_{12}}+\frac{(d-4)p_{3}^{\mu_{2}}p_{2}^{\mu_{1}}}{s_{23}}+\frac{(d-2)s_{13}p_{2}^{\mu_{2}}p_{2}^{\mu_{1}}}{s_{12}s_{23}}+\frac{(d-2)s_{23}p_{1}^{\mu_{1}}p_{1}^{\mu_{2}}}{s_{12}s_{13}}\nonumber \\
 & \qquad-\frac{(d-2)p_{1}^{\mu_{1}}p_{2}^{\mu_{2}}}{s_{12}}-\frac{(d-2)p_{1}^{\mu_{2}}p_{3}^{\mu_{1}}}{s_{13}}-\frac{(d-2)p_{1}^{\mu_{1}}p_{3}^{\mu_{2}}}{s_{13}}+\frac{(d-4)p_{2}^{\mu_{2}}p_{3}^{\mu_{1}}}{s_{23}}+\frac{(d-2)s_{12}p_{3}^{\mu_{1}}p_{3}^{\mu_{2}}}{s_{13}s_{23}}\Bigg]\,,\\
P_{5}^{\mu_{1}\mu_{2}} & =\frac{1}{\left(d-3\right)}\Bigg[g^{\mu_{1}\mu_{2}}-\frac{(d-2)p_{1}^{\mu_{2}}p_{2}^{\mu_{1}}}{s_{12}}-\frac{(d-2)p_{3}^{\mu_{2}}p_{2}^{\mu_{1}}}{s_{23}}+\frac{(d-2)s_{13}p_{2}^{\mu_{2}}p_{2}^{\mu_{1}}}{s_{12}s_{23}}+\frac{(d-2)s_{23}p_{1}^{\mu_{1}}p_{1}^{\mu_{2}}}{s_{12}s_{13}}\nonumber \\
 & \qquad-\frac{(d-2)p_{1}^{\mu_{1}}p_{2}^{\mu_{2}}}{s_{12}}+\frac{(d-4)p_{1}^{\mu_{2}}p_{3}^{\mu_{1}}}{s_{13}}+\frac{(d-4)p_{1}^{\mu_{1}}p_{3}^{\mu_{2}}}{s_{13}}-\frac{(d-2)p_{2}^{\mu_{2}}p_{3}^{\mu_{1}}}{s_{23}}+\frac{(d-2)s_{12}p_{3}^{\mu_{1}}p_{3}^{\mu_{2}}}{s_{13}s_{23}}\Bigg]\,.
\end{align}
\label{eq:projs}
\end{subequations}

%% file: main.bbl
\providecommand{\href}[2]{#2}\begingroup\raggedright\begin{thebibliography}{10}

\bibitem{Cvitanovic:1980bu}
P.~Cvitanovic, P.~G. Lauwers and P.~N. Scharbach, \emph{{Gauge Invariance
  Structure of Quantum Chromodynamics}},
  \href{http://dx.doi.org/10.1016/0550-3213(81)90098-5}{\emph{Nucl. Phys.} {\bf
  B186} (1981) 165--186}.

\bibitem{Berends:1987cv}
F.~A. Berends and W.~Giele, \emph{{The Six Gluon Process as an Example of
  Weyl-Van Der Waerden Spinor Calculus}},
  \href{http://dx.doi.org/10.1016/0550-3213(87)90604-3}{\emph{Nucl. Phys.} {\bf
  B294} (1987) 700--732}.

\bibitem{Mangano:1987xk}
M.~L. Mangano, S.~J. Parke and Z.~Xu, \emph{{Duality and Multi - Gluon
  Scattering}},
  \href{http://dx.doi.org/10.1016/0550-3213(88)90001-6}{\emph{Nucl.Phys.} {\bf
  B298} (1988) 653}.

\bibitem{Kosower:1987ic}
D.~Kosower, B.-H. Lee and V.~P. Nair, \emph{{MULTI GLUON SCATTERING: A STRING
  BASED CALCULATION}},
  \href{http://dx.doi.org/10.1016/0370-2693(88)90085-8}{\emph{Phys. Lett.} {\bf
  B201} (1988) 85--89}.

\bibitem{Bern:1990ux}
Z.~Bern and D.~A. Kosower, \emph{{Color decomposition of one loop amplitudes in
  gauge theories}},
  \href{http://dx.doi.org/10.1016/0550-3213(91)90567-H}{\emph{Nucl. Phys.} {\bf
  B362} (1991) 389--448}.

\bibitem{DelDuca:1999rs}
V.~Del~Duca, L.~J. Dixon and F.~Maltoni, \emph{{New color decompositions for
  gauge amplitudes at tree and loop level}},
  \href{http://dx.doi.org/10.1016/S0550-3213(99)00809-3}{\emph{Nucl. Phys.}
  {\bf B571} (2000) 51--70}, [\href{http://arxiv.org/abs/hep-ph/9910563}{{\tt
  hep-ph/9910563}}].

\bibitem{Maltoni:2002mq}
F.~Maltoni, K.~Paul, T.~Stelzer and S.~Willenbrock, \emph{{Color flow
  decomposition of QCD amplitudes}},
  \href{http://dx.doi.org/10.1103/PhysRevD.67.014026}{\emph{Phys. Rev.} {\bf
  D67} (2003) 014026}, [\href{http://arxiv.org/abs/hep-ph/0209271}{{\tt
  hep-ph/0209271}}].

\bibitem{Kleiss:1988ne}
R.~Kleiss and H.~Kuijf, \emph{{Multi - Gluon Cross-sections and Five Jet
  Production at Hadron Colliders}},
  \href{http://dx.doi.org/10.1016/0550-3213(89)90574-9}{\emph{Nucl. Phys.} {\bf
  B312} (1989) 616}.

\bibitem{Bern:2008qj}
Z.~Bern, J.~Carrasco and H.~Johansson, \emph{{New Relations for Gauge-Theory
  Amplitudes}},
  \href{http://dx.doi.org/10.1103/PhysRevD.78.085011}{\emph{Phys.Rev.} {\bf
  D78} (2008) 085011}, [\href{http://arxiv.org/abs/0805.3993}{{\tt
  0805.3993}}].

\bibitem{Bern:2010ue}
Z.~Bern, J.~J.~M. Carrasco and H.~Johansson, \emph{{Perturbative Quantum
  Gravity as a Double Copy of Gauge Theory}},
  \href{http://dx.doi.org/10.1103/PhysRevLett.105.061602}{\emph{Phys.Rev.Lett.}
  {\bf 105} (2010) 061602}, [\href{http://arxiv.org/abs/1004.0476}{{\tt
  1004.0476}}].

\bibitem{Bern:2012uf}
Z.~Bern, J.~Carrasco, L.~Dixon, H.~Johansson and R.~Roiban, \emph{{Simplifying
  Multiloop Integrands and Ultraviolet Divergences of Gauge Theory and Gravity
  Amplitudes}},
  \href{http://dx.doi.org/10.1103/PhysRevD.85.105014}{\emph{Phys.Rev.} {\bf
  D85} (2012) 105014}, [\href{http://arxiv.org/abs/1201.5366}{{\tt
  1201.5366}}].

\bibitem{Carrasco:2011mn}
J.~J. Carrasco and H.~Johansson, \emph{{Five-Point Amplitudes in N=4
  Super-Yang-Mills Theory and N=8 Supergravity}},
  \href{http://dx.doi.org/10.1103/PhysRevD.85.025006}{\emph{Phys.Rev.} {\bf
  D85} (2012) 025006}, [\href{http://arxiv.org/abs/1106.4711}{{\tt
  1106.4711}}].

\bibitem{Bjerrum-Bohr:2013iza}
N.~E.~J. Bjerrum-Bohr, T.~Dennen, R.~Monteiro and D.~O'Connell,
  \emph{{Integrand Oxidation and One-Loop Colour-Dual Numerators in N=4 Gauge
  Theory}}, \href{http://dx.doi.org/10.1007/JHEP07(2013)092}{\emph{JHEP} {\bf
  1307} (2013) 092}, [\href{http://arxiv.org/abs/1303.2913}{{\tt 1303.2913}}].

\bibitem{Carrasco:2012ca}
J.~J.~M. Carrasco, M.~Chiodaroli, M.~G{\"u}naydin and R.~Roiban,
  \emph{{One-loop four-point amplitudes in pure and matter-coupled N <= 4
  supergravity}}, \href{http://dx.doi.org/10.1007/JHEP03(2013)056}{\emph{JHEP}
  {\bf 1303} (2013) 056}, [\href{http://arxiv.org/abs/1212.1146}{{\tt
  1212.1146}}].

\bibitem{Bern:2011rj}
Z.~Bern, C.~Boucher-Veronneau and H.~Johansson, \emph{{N >= 4 Supergravity
  Amplitudes from Gauge Theory at One Loop}},
  \href{http://dx.doi.org/10.1103/PhysRevD.84.105035}{\emph{Phys.Rev.} {\bf
  D84} (2011) 105035}, [\href{http://arxiv.org/abs/1107.1935}{{\tt
  1107.1935}}].

\bibitem{BoucherVeronneau:2011qv}
C.~Boucher-Veronneau and L.~Dixon, \emph{{N >- 4 Supergravity Amplitudes from
  Gauge Theory at Two Loops}},
  \href{http://dx.doi.org/10.1007/JHEP12(2011)046}{\emph{JHEP} {\bf 1112}
  (2011) 046}, [\href{http://arxiv.org/abs/1110.1132}{{\tt 1110.1132}}].

\bibitem{Mastrolia:2012wf}
P.~Mastrolia, E.~Mirabella, G.~Ossola and T.~Peraro, \emph{{Integrand-Reduction
  for Two-Loop Scattering Amplitudes through Multivariate Polynomial
  Division}}, \href{http://dx.doi.org/10.1103/PhysRevD.87.085026}{\emph{Phys.
  Rev.} {\bf D87} (2013) 085026}, [\href{http://arxiv.org/abs/1209.4319}{{\tt
  1209.4319}}].

\bibitem{Yang:2016ear}
G.~Yang, \emph{{Color-kinematics duality and Sudakov form factor at five loops
  for N=4 supersymmetric Yang-Mills theory}},
  \href{http://dx.doi.org/10.1103/PhysRevLett.117.271602}{\emph{Phys. Rev.
  Lett.} {\bf 117} (2016) 271602}, [\href{http://arxiv.org/abs/1610.02394}{{\tt
  1610.02394}}].

\bibitem{Bern:2017ucb}
Z.~Bern, J.~J.~M. Carrasco, W.-M. Chen, H.~Johansson, R.~Roiban and M.~Zeng,
  \emph{{The Five-Loop Four-Point Integrand of N=8 Supergravity as a
  Generalized Double Copy}},  \href{http://arxiv.org/abs/1708.06807}{{\tt
  1708.06807}}.

\bibitem{Bern:2013yya}
Z.~Bern, S.~Davies, T.~Dennen, Y.-t. Huang and J.~Nohle,
  \emph{{Color-Kinematics Duality for Pure Yang-Mills and Gravity at One and
  Two Loops}}, \href{http://dx.doi.org/10.1103/PhysRevD.92.045041}{\emph{Phys.
  Rev.} {\bf D92} (2015) 045041}, [\href{http://arxiv.org/abs/1303.6605}{{\tt
  1303.6605}}].

\bibitem{Nohle:2013bfa}
J.~Nohle, \emph{{Color-Kinematics Duality in One-Loop Four-Gluon Amplitudes
  with Matter}},
  \href{http://dx.doi.org/10.1103/PhysRevD.90.025020}{\emph{Phys.Rev.} {\bf
  D90} (2014) 025020}, [\href{http://arxiv.org/abs/1309.7416}{{\tt
  1309.7416}}].

\bibitem{Badger:2015lda}
S.~Badger, G.~Mogull, A.~Ochirov and D.~O'Connell, \emph{{A Complete Two-Loop,
  Five-Gluon Helicity Amplitude in Yang-Mills Theory}},
  \href{http://dx.doi.org/10.1007/JHEP10(2015)064}{\emph{JHEP} {\bf 10} (2015)
  064}, [\href{http://arxiv.org/abs/1507.08797}{{\tt 1507.08797}}].

\bibitem{BjerrumBohr:2010zs}
N.~Bjerrum-Bohr, P.~H. Damgaard, T.~Sondergaard and P.~Vanhove,
  \emph{{Monodromy and Jacobi-like Relations for Color-Ordered Amplitudes}},
  \href{http://dx.doi.org/10.1007/JHEP06(2010)003}{\emph{JHEP} {\bf 1006}
  (2010) 003}, [\href{http://arxiv.org/abs/1003.2403}{{\tt 1003.2403}}].

\bibitem{Tourkine:2016bak}
P.~Tourkine and P.~Vanhove, \emph{{Higher-loop amplitude monodromy relations in
  string and gauge theory}},
  \href{http://dx.doi.org/10.1103/PhysRevLett.117.211601}{\emph{Phys. Rev.
  Lett.} {\bf 117} (2016) 211601}, [\href{http://arxiv.org/abs/1608.01665}{{\tt
  1608.01665}}].

\bibitem{Ochirov:2017jby}
A.~Ochirov, P.~Tourkine and P.~Vanhove, \emph{{One-loop monodromy relations on
  single cuts}},  \href{http://arxiv.org/abs/1707.05775}{{\tt 1707.05775}}.

\bibitem{He:2017spx}
S.~He, O.~Schlotterer and Y.~Zhang, \emph{{New BCJ representations for one-loop
  amplitudes in gauge theories and gravity}},
  \href{http://arxiv.org/abs/1706.00640}{{\tt 1706.00640}}.

\bibitem{Chester:2016ojq}
D.~Chester, \emph{{Bern-Carrasco-Johansson relations for one-loop QCD integral
  coefficients}},
  \href{http://dx.doi.org/10.1103/PhysRevD.93.065047}{\emph{Phys. Rev.} {\bf
  D93} (2016) 065047}, [\href{http://arxiv.org/abs/1601.00235}{{\tt
  1601.00235}}].

\bibitem{Primo:2016omk}
A.~Primo and W.~J. Torres~Bobadilla, \emph{{BCJ Identities and $d$-Dimensional
  Generalized Unitarity}},
  \href{http://dx.doi.org/10.1007/JHEP04(2016)125}{\emph{JHEP} {\bf 04} (2016)
  125}, [\href{http://arxiv.org/abs/1602.03161}{{\tt 1602.03161}}].

\bibitem{Mastrolia:2015maa}
P.~Mastrolia, A.~Primo, U.~Schubert and W.~J. Torres~Bobadilla,
  \emph{{Off-shell currents and color--kinematics duality}},
  \href{http://dx.doi.org/10.1016/j.physletb.2015.11.084}{\emph{Phys. Lett.}
  {\bf B753} (2016) 242--262}, [\href{http://arxiv.org/abs/1507.07532}{{\tt
  1507.07532}}].

\bibitem{Bern:2010yg}
Z.~Bern, T.~Dennen, Y.-t. Huang and M.~Kiermaier, \emph{{Gravity as the Square
  of Gauge Theory}},
  \href{http://dx.doi.org/10.1103/PhysRevD.82.065003}{\emph{Phys.Rev.} {\bf
  D82} (2010) 065003}, [\href{http://arxiv.org/abs/1004.0693}{{\tt
  1004.0693}}].

\bibitem{BjerrumBohr:2012mg}
N.~Bjerrum-Bohr, P.~H. Damgaard, R.~Monteiro and D.~O'Connell, \emph{{Algebras
  for Amplitudes}},
  \href{http://dx.doi.org/10.1007/JHEP06(2012)061}{\emph{JHEP} {\bf 1206}
  (2012) 061}, [\href{http://arxiv.org/abs/1203.0944}{{\tt 1203.0944}}].

\bibitem{Boels:2012sy}
R.~H. Boels and R.~S. Isermann, \emph{{On powercounting in perturbative quantum
  gravity theories through color-kinematic duality}},
  \href{http://dx.doi.org/10.1007/JHEP06(2013)017}{\emph{JHEP} {\bf 06} (2013)
  017}, [\href{http://arxiv.org/abs/1212.3473}{{\tt 1212.3473}}].

\bibitem{Catani:2008xa}
S.~Catani, T.~Gleisberg, F.~Krauss, G.~Rodrigo and J.-C. Winter, \emph{{From
  loops to trees by-passing Feynman's theorem}},
  \href{http://dx.doi.org/10.1088/1126-6708/2008/09/065}{\emph{JHEP} {\bf 09}
  (2008) 065}, [\href{http://arxiv.org/abs/0804.3170}{{\tt 0804.3170}}].

\bibitem{Sborlini:2016hat}
G.~F.~R. Sborlini, F.~Driencourt-Mangin and G.~Rodrigo, \emph{{Four-dimensional
  unsubtraction with massive particles}},
  \href{http://dx.doi.org/10.1007/JHEP10(2016)162}{\emph{JHEP} {\bf 10} (2016)
  162}, [\href{http://arxiv.org/abs/1608.01584}{{\tt 1608.01584}}].

\bibitem{Sborlini:2016gbr}
G.~F.~R. Sborlini, F.~Driencourt-Mangin, R.~Hernandez-Pinto and G.~Rodrigo,
  \emph{{Four-dimensional unsubtraction from the loop-tree duality}},
  \href{http://dx.doi.org/10.1007/JHEP08(2016)160}{\emph{JHEP} {\bf 08} (2016)
  160}, [\href{http://arxiv.org/abs/1604.06699}{{\tt 1604.06699}}].

\bibitem{Hernandez-Pinto:2015ysa}
R.~J. Hernandez-Pinto, G.~F.~R. Sborlini and G.~Rodrigo, \emph{{Towards gauge
  theories in four dimensions}},
  \href{http://dx.doi.org/10.1007/JHEP02(2016)044}{\emph{JHEP} {\bf 02} (2016)
  044}, [\href{http://arxiv.org/abs/1506.04617}{{\tt 1506.04617}}].

\bibitem{Buchta:2015wna}
S.~Buchta, G.~Chachamis, P.~Draggiotis and G.~Rodrigo, \emph{{Numerical
  implementation of the loop--tree duality method}},
  \href{http://dx.doi.org/10.1140/epjc/s10052-017-4833-6}{\emph{Eur. Phys. J.}
  {\bf C77} (2017) 274}, [\href{http://arxiv.org/abs/1510.00187}{{\tt
  1510.00187}}].

\bibitem{Hahn:2000kx}
T.~Hahn, \emph{{Generating Feynman diagrams and amplitudes with FeynArts 3}},
  \href{http://dx.doi.org/10.1016/S0010-4655(01)00290-9}{\emph{Comput. Phys.
  Commun.} {\bf 140} (2001) 418--431},
  [\href{http://arxiv.org/abs/hep-ph/0012260}{{\tt hep-ph/0012260}}].

\bibitem{Mertig:1990an}
R.~Mertig, M.~Bohm and A.~Denner, \emph{{FEYN CALC: Computer algebraic
  calculation of Feynman amplitudes}},
  \href{http://dx.doi.org/10.1016/0010-4655(91)90130-D}{\emph{Comput. Phys.
  Commun.} {\bf 64} (1991) 345--359}.

\bibitem{Shtabovenko:2016sxi}
V.~Shtabovenko, R.~Mertig and F.~Orellana, \emph{{New Developments in FeynCalc
  9.0}},  \href{http://arxiv.org/abs/1601.01167}{{\tt 1601.01167}}.

\bibitem{Bern:2010tq}
Z.~Bern, J.~Carrasco, L.~J. Dixon, H.~Johansson and R.~Roiban, \emph{{The
  Complete Four-Loop Four-Point Amplitude in N=4 Super-Yang-Mills Theory}},
  \href{http://dx.doi.org/10.1103/PhysRevD.82.125040}{\emph{Phys.Rev.} {\bf
  D82} (2010) 125040}, [\href{http://arxiv.org/abs/1008.3327}{{\tt
  1008.3327}}].

\bibitem{Tolotti:2013caa}
M.~Tolotti and S.~Weinzierl, \emph{{Construction of an effective Yang-Mills
  Lagrangian with manifest BCJ duality}},
  \href{http://dx.doi.org/10.1007/JHEP07(2013)111}{\emph{JHEP} {\bf 1307}
  (2013) 111}, [\href{http://arxiv.org/abs/1306.2975}{{\tt 1306.2975}}].

\bibitem{Melia:2013bta}
T.~Melia, \emph{{Dyck words and multiquark primitive amplitudes}},
  \href{http://dx.doi.org/10.1103/PhysRevD.88.014020}{\emph{Phys. Rev.} {\bf
  D88} (2013) 014020}, [\href{http://arxiv.org/abs/1304.7809}{{\tt
  1304.7809}}].

\bibitem{Melia:2013epa}
T.~Melia, \emph{{Getting more flavor out of one-flavor QCD}},
  \href{http://dx.doi.org/10.1103/PhysRevD.89.074012}{\emph{Phys. Rev.} {\bf
  D89} (2014) 074012}, [\href{http://arxiv.org/abs/1312.0599}{{\tt
  1312.0599}}].

\bibitem{Johansson:2014zca}
H.~Johansson and A.~Ochirov, \emph{{Pure Gravities via Color-Kinematics Duality
  for Fundamental Matter}},
  \href{http://dx.doi.org/10.1007/JHEP11(2015)046}{\emph{JHEP} {\bf 11} (2015)
  046}, [\href{http://arxiv.org/abs/1407.4772}{{\tt 1407.4772}}].

\bibitem{Johansson:2015oia}
H.~Johansson and A.~Ochirov, \emph{{Color-Kinematics Duality for QCD
  Amplitudes}}, \href{http://dx.doi.org/10.1007/JHEP01(2016)170}{\emph{JHEP}
  {\bf 01} (2016) 170}, [\href{http://arxiv.org/abs/1507.00332}{{\tt
  1507.00332}}].

\bibitem{Driencourt-Mangin:2017gop}
F.~Driencourt-Mangin, G.~Rodrigo and G.~F.~R. Sborlini, \emph{{Universal dual
  amplitudes and asymptotic expansions for $gg\to H$ and $H\to \gamma\gamma$ in
  four dimensions}},  \href{http://arxiv.org/abs/1702.07581}{{\tt 1702.07581}}.

\bibitem{Tkachov:1981wb}
F.~V. Tkachov, \emph{{A Theorem on Analytical Calculability of Four Loop
  Renormalization Group Functions}},
  \href{http://dx.doi.org/10.1016/0370-2693(81)90288-4}{\emph{Phys. Lett.} {\bf
  B100} (1981) 65--68}.

\bibitem{Chetyrkin:1981qh}
K.~G. Chetyrkin and F.~V. Tkachov, \emph{{Integration by Parts: The Algorithm
  to Calculate beta Functions in 4 Loops}},
  \href{http://dx.doi.org/10.1016/0550-3213(81)90199-1}{\emph{Nucl. Phys.} {\bf
  B192} (1981) 159--204}.

\bibitem{Laporta:2001dd}
S.~Laporta, \emph{{High precision calculation of multiloop Feynman integrals by
  difference equations}},
  \href{http://dx.doi.org/10.1016/S0217-751X(00)00215-7,
  10.1142/S0217751X00002157}{\emph{Int. J. Mod. Phys.} {\bf A15} (2000)
  5087--5159}, [\href{http://arxiv.org/abs/hep-ph/0102033}{{\tt
  hep-ph/0102033}}].

\end{thebibliography}\endgroup
